\documentclass{emulateapj}

\shorttitle{A Study of Edge-On Galaxies with HST/ACS. I.}
\shortauthors{Seth, Dalcanton \& de Jong}

\begin{document}

\slugcomment{Re-Submitted Version, Nov. 8th, 2004}

\title{A Study of Edge-On Galaxies with the Hubble Space Telescope's
  Advanced Camera for Surveys. I. Initial Results.}

\author{Anil C. Seth}
\affil{University of Washington}
\email{seth@astro.washington.edu}

\author{Julianne J. Dalcanton\footnote{Alfred P. Sloan Research Fellow}}
\affil{University of Washington}
\email{jd@astro.washington.edu}

\author{Roelof S. de Jong}
\affil{Space Telescope Science Institute}
\email{dejong@stsci.edu}

\begin{abstract} 
We present the initial results of a Hubble Space Telescope/ACS
snapshot survey of 16 nearby, edge-on, late-type galaxies covering a
range in distance from 2 to 19 Mpc.  The 
images of these galaxies show significant resolved
stellar populations.  We derive F606W and F814W
photometry for $>$1.2 million stars, and present 
color-magnitude diagrams which show a mixture of young,
intermediate, and old stars in each galaxy.  In one of the fields we
serendipitously 
detect stars from the Large Magellanic Cloud.  We also identify a
candidate young dwarf galaxy lying $\sim$2 kpc above the plane of NGC 4631.   
For the nearest six galaxies, we derive tip of the red giant
branch distances and demonstrate that these galaxies fall on the
K-band Tully-Fisher relation established in clusters.  From the color
of the red giant branch, we also find evidence that these galaxies
possess a metal-poor thick disk or halo population.  

\end{abstract}

\section{Introduction}

This paper is the first in a series presenting results from a
Hubble Space Telescope (HST) snapshot program of nearby edge-on galaxies.
These galaxies are mostly low mass, late-type spirals with circular
velocities below 200 km/sec.  
Although the project was designed to study the dust
content of the galaxies, a 
number of the galaxies were close enough to resolve a significant
population of stars and it is the resolved stars that we focus on
here.  Previous studies of resolved stellar populations with HST have
shown that they can be used to answer a variety of questions.
The ages and metallicities of the stars can be used to address issues
of galaxy formation
\citep[e.g.][]{butler04,brown03,dolphin03,sarajedini01} and distances
can be found using the tip of the red giant branch (TRGB) method
\citep[e.g.][]{mendez02,karachentsev03a}.
In this paper we present the galaxies' images, stellar photometry, and some
initial results, including TRGB distance measurements, the
identification of a 
possible young dwarf galaxy near NGC 4631, and the detection of LMC
stars $\sim$9$^\circ$ away from the galaxy center.  In the second
paper 
\citep[Paper II,][]{seth04} we analyze the variation in the vertical
distribution of different stellar populations and present evidence
that many of these galaxies host metal-poor thick disks.  

First, in \S2 we describe our sample, present
a detailed description of our image and 
photometry data reduction process, and discuss the stellar models
and photometric transformations we use to analyze our data.  In
\S3 we present images 
and color magnitude diagrams of each galaxy and identify the visible
stellar populations.  We also identify a candidate dwarf galaxy lying just
off the plane of NGC 4631, which appears to contain only young stars.
Our method and results for TRGB distance
determination are discussed in \S4 and conclusions follow in \S5.

\section{Sample Description \& Data Reduction}

Our sample consists of 16 galaxies observed with the
Advanced Camera for Surveys (ACS) as part of an HST Cycle 12 snapshot project
(PI: de Jong, PID: 9765).  The sample was selected from the
\citet{tully88} and \citet{karachentsev93} catalogs to be edge-on and
nearby.  Of the 40 fields submitted for observation, 18 fields were
observed in 16 separate galaxies.  

\begin{deluxetable*}{lcccccccccc}
\tablecaption{Galaxy Sample Properties}
\tabletypesize{\scriptsize}
\tablehead{
     \colhead{Galaxy}  &
     \colhead{RA(J2000)} &
     \colhead{Dec(J2000)} &
     \colhead{Pos. Ang.} &
     \colhead{b [deg]\tablenotemark{a}} &
     \colhead{Type} &
     \colhead{B$_{\rm T}$} &
     \colhead{K$_{\rm tot}$\tablenotemark{b}} &
     \colhead{V$_{\rm max}$} &
     \colhead{V$_{\rm recess}$} &
     \colhead{m-M$_{\rm lit}$}
}

\startdata
IC 2233         & 08 13 58.8  & +45 44 31 &  -7.6 &  33.06 & SBcd  & 13.36 & 10.75 & 79  &  558 & 29.96\tablenotemark{f} \\
IC 5052         & 20 52 02.9  & -69 11 45 & -38.0 & -35.80 & SBcd  & 11.67 &  8.88 & 79  &  598 & 28.80\tablenotemark{g} \\ 
IRAS 06070-6147\tablenotemark{c} & 06 07 29.9  & -61 48 26 &  42.7 & -28.82 & Sc    & 10.72 &  8.98 & 133 & 1227 & 31.22\tablenotemark{h} \\ 
IRAS 07568-4942\tablenotemark{d} & 07 58 15.3  & -49 51 13 & -26.5 & -10.58 & SBc   & 12.36 &  8.04 & 142 & 1119 & 30.06\tablenotemark{g} \\ 
IRAS 09312-3248\tablenotemark{e} & 09 33 21.2  & -33 02 00 &  89.7 &  13.64 & Sc    & 12.71 &  8.88 & 98  &  929 & 30.61\tablenotemark{h} \\ 
NGC 55          & 00 14 54.4  & -39 11 59 & -70.0 & -75.73 & SBm   & 9.58  &  6.25 & 67  &  131 & 26.28\tablenotemark{i} \\ 
NGC 891         & 02 22 33.0  & +42 20 53 &  22.8 & -17.41 & Sb    & 10.83 &  5.94 & 214 &  530 & 29.61\tablenotemark{j} \\ 
NGC 3501        & 11 02 47.3  & +17 59 22 &  28.1 &  63.37 & Sc    & 13.67 &  9.41 & 136 & 1132 & 31.04\tablenotemark{h} \\ 
NGC 4144        & 12 09 58.3  & +46 27 28 & -77.6 &  69.01 & SBc   & 12.16 &  9.39 & 67  &  267 & 27.93\tablenotemark{g} \\ 
NGC 4183        & 12 13 16.8  & +43 41 53 & -14.0 &  71.73 & Sc    & 12.94 &  9.85 & 103 &  931 & 31.35\tablenotemark{k} \\ 
NGC 4206        & 12 15 16.8  & +13 01 27 &   1.0 &  73.55 & Sbc   & 12.83 &  9.39 & 124 &  712 & 31.29\tablenotemark{l} \\ 
NGC 4244        & 12 17 29.5  & +37 48 28 &  47.2 &  77.16 & Sc    & 11.12 &  7.72 & 93  &  244 & 28.26\tablenotemark{m} \\ 
NGC 4517        & 12 32 45.6  & +00 06 56 &  82.6 &  62.62 & Sc    & 11.12 &  7.33 & 137 & 1126 & 29.30\tablenotemark{f} \\ 
NGC 4565        & 12 36 20.7  & +25 59 15 & -44.8 &  86.44 & Sb    & 10.58 &  6.06 & 227 & 1228 & 31.05\tablenotemark{m} \\ 
NGC 4631        & 12 42 07.7  & +32 32 33 &  82.6 &  84.22 & SBcd  & 9.67  &  6.47 & 131 &  599 & 28.18\tablenotemark{f} \\ 
NGC 5023        & 13 12 11.7  & +44 02 17 &  27.9 &  72.58 & Sc    & 12.81 &  8.74 & 77  &  406 & 28.75\tablenotemark{f} \\ 
\enddata
\tablecomments{RA, Dec and Position Angle from K band galaxy fits
  described in Appendix.  All other data from HYPERLEDA/LEDA
  \citep{paturel95,paturel03} except where noted. 
  (a) Galactic latitude, (b) Total K-band magnitude from \citet{jarrett03},
  (c) IRAS~06070-6147 also known as ESO~121-G6,
  (d) IRAS~07568-4942 also known as ESO~209-G9,
  (e) IRAS~06070-6147 also known as ESO~373-G8 or UGCA~168,
  (f) \citet{bottinelli85}, (g) \citet{bottinelli88}, (h)
  assuming H$_0$=70 km/sec, (i) \citet{karachentsev03a}, (j)
  \citet{tonry01}, (k) \citet{tully00}, (l) \citet{gavazzi99},
  (m) \citet{karachentsev03b}, (n) \citet{jensen03}
}

\end{deluxetable*}

The properties of each galaxy in the sample is given
in Table~1.  With the exception of NGC 891 \& NGC 4565, the galaxies
are all later than Sb and have maximum circular velocities below 200
km/sec.  Thus, most have a significantly lower mass than the
Milky Way.  Based on previously published distances (m-M$_{\rm lit}$
in Table~1), mostly determined
using the Tully-Fisher method, the nearest of the galaxies is NGC 55
(1.8 Mpc \citep{karachentsev03a}) and the most distant is NGC 4183 (19
Mpc \citep{tully00}).  Assuming these distances, the absolute
B band magnitudes of the galaxies range from roughly -16 for NGC 4144 \& NGC 5023 to
roughly -20.5 for IRAS 06070-6147 \& NGC 4565.  For comparison, the
Large Magellanic Cloud has a absolute B magnitude of -17.6, using the
B$_{\rm T}$ magnitude from the LEDA database \citep{paturel95} and
assuming a distance modulus of 18.50 \citep{alves04}.  We also
performed model fits to 2MASS K band images to obtain a consistent set
of structural properties for use in our analysis.  These fits are
described in detail in the 
Appendix.  In Table~2 we show the resulting scale length ($h_R$),
scale height (z$_0$), and half-light height z$_{1/2}$.  Note that for
the model used (Eq.~1 of the Appendix), z$_{1/2}$=0.549z$_0$.

\begin{deluxetable*}{lccccccc}
\tabletypesize{\scriptsize}
\tablecaption{K Band disk fit parameters}
\tablehead{
     \colhead{Galaxy}  &
     \colhead{$\Sigma_0$} &
     \colhead{$h_{R}$} &
     \colhead{z$_0$} & 
     \colhead{$h_{R}$/z$_0$} & 
     \colhead{z$_{1/2}$} &
     \colhead{$h_{R}$} &
     \colhead{z$_0$} \\
     \colhead{}  &
     \colhead{[mag/$\square$\arcsec]} &
     \colhead{[\arcsec]} &
     \colhead{[\arcsec]} & 
     \colhead{} & 
     \colhead{[\arcsec]} &
     \colhead{[pc]\tablenotemark{a}} &
     \colhead{[pc]\tablenotemark{a}} 
}

\startdata
IC 2233         &  18.42  &  25.97  &   6.87 & 3.78 &  3.77 & 1236 &  327 \\
IC 5052         &  18.02  &  53.87  &  13.35 & 4.04 &  7.33 & 1574 &  390 \\ 
IRAS 06070-6147 &  15.92  &  16.36  &   4.07 & 4.02 &  2.23 & 1391 &  346 \\ 
IRAS 07568-4942 &  16.05  &  27.80  &   6.87 & 4.05 &  3.77 & 1386 &  342 \\ 
IRAS 09312-3248 &  17.52  &  38.24  &  10.33 & 3.70 &  5.67 & 2455 &  663 \\ 
NGC 55          &  17.38  &  93.30  &  42.52 & 2.19 & 23.34 &  958 &  437 \\ 
NGC 891         &  15.59  &  90.75  &  11.86 & 7.65 &  6.51 & 3676 &  481 \\ 
NGC 3501        &  16.20  &  19.96  &   4.15 & 4.81 &  2.28 & 1562 &  325 \\ 
NGC 4144        &  18.01  &  30.56  &  12.68 & 2.41 &  6.96 & 1103 &  458 \\ 
NGC 4183        &  17.47  &  19.66  &   7.21 & 2.73 &  3.96 & 1775 &  651 \\ 
NGC 4206        &  17.24  &  24.44  &   7.45 & 3.28 &  4.09 & 2146 &  654 \\ 
NGC 4244        &  17.91  &  84.27  &  22.14 & 3.81 & 12.15 & 1783 &  469 \\ 
NGC 4517        &  17.02  &  79.13  &  14.64 & 5.41 &  8.04 & 2779 &  514 \\ 
NGC 4565        &  16.02  &  87.86  &  13.67 & 6.43 &  7.50 & 6908 & 1075 \\ 
NGC 4631        &  15.62  &  35.49  &  13.74 & 2.58 &  7.54 & 1323 &  512 \\ 
NGC 5023        &  18.16  &  39.81  &   9.29 & 4.29 &  5.10 & 1275 &  298 \\ 
\enddata
\tablecomments{(a) Derived using distances from Table~1 or when
  available, Table~5} 
\end{deluxetable*}

The HST imaging was obtained using the Wide
Field Camera (WFC) camera of ACS.  A total of 18 fields in
16 galaxies were observed with the F606W and F814W filters (see
Table~3).  Each field was observed with one of two observing plans:   
\begin{itemize}
\item {\bf 2$\times$2:} 14 fields have two exposures of 338
  seconds in F606W and two exposures of 350 seconds in F814W.
\item {\bf 1$\times$2:} 4 fields have one 400 second exposure in F606W
  and two 338 second exposures in F814W.
\end{itemize}

\begin{deluxetable}{lcr}
\tabletypesize{\scriptsize}
\tablecaption{Observed HST Fields}
\tablehead{
     \colhead{Field}  &
     \colhead{Obs. Plan\tablenotemark{a}} &
     \colhead{\# Stars}
}

\startdata
IC 2233         & 2$\times$2 &  15697 \\
IC 5052         & 2$\times$2 &  68636 \\ 
IRAS 06070-6147 & 1$\times$2 &  15271 \\ 
IRAS 07568-4942 & 1$\times$2 &  23582 \\ 
IRAS 09312-3248 & 2$\times$2 &  21667 \\ 
NGC 55          & 1$\times$2 & 281536 \\ 
NGC 55-DISK     & 2$\times$2 & 253108 \\ 
NGC 891         & 2$\times$2 &  33508 \\ 
NGC 3501        & 2$\times$2 &   4887 \\ 
NGC 4144        & 2$\times$2 &  60552 \\ 
NGC 4183        & 2$\times$2 &  13103 \\ 
NGC 4206        & 2$\times$2 &   8733 \\ 
NGC 4244        & 2$\times$2 & 121238 \\ 
NGC 4517        & 2$\times$2 &  49201 \\ 
NGC 4565        & 2$\times$2 &  20192 \\ 
NGC 4631        & 2$\times$2 & 104940 \\ 
NGC 4631-DISK   & 2$\times$2 &  97656 \\ 
NGC 5023        & 1$\times$2 &  42293 \\ 
\enddata
\tablecomments{{\footnotesize   
{\it (a) }{{\bf 2$\times$2:} denotes fields with two 338 second
  exposures in F606W and two 350 second exposures in
  F814W. {\bf 1$\times$2:} denotes fields with one 400 second exposure
  in F606W and two 338 second exposures in F814W.}
}}
\end{deluxetable}

\noindent In all cases where two exposures were taken, they were
dithered to fill in the gap between the two ACS/WFC CCDs.   

We developed the following data reduction pipeline for these galaxies:

\begin{enumerate}
\item Removal of cosmic rays using the Laplacian edge-detection
  routine of \citet{vandokkum01} on the raw FLT images obtained from
  the HST/MAST archive.
\item Sky subtraction, drizzling and further cosmic ray removal using
  the PyRAF 'multidrizzle' task (code written by Anton Koekemoer).
\item Location and aperture photometry of objects in the image using
  DAOPHOT II \citep{stetson87}.
\item Determination of the point-spread function (PSF) and PSF
  photometry also using DAOPHOT II.
\item Artificial star tests to obtain the completeness as a
  function of position in our images.
\end{enumerate}

All 18 fields were reduced using this same pipeline.
Noting the lack of detailed descriptions of ACS data reduction and
photometry in the literature, we will now describe each of these steps
in detail.  

\subsection{Cosmic Ray Removal \& Drizzling}

Based on ACS report 99-09 (Mutchler et al.), we expect
to find that approximately 1.5\% of pixels ($\sim$300,000 pixels in
our final 4350$\times$4350 images) are affected by cosmic
rays.  Because our data set includes only two images, we hoped to be
able to reject cosmic rays in each single image before combination.

Tests of the Laplacian edge detection method on WFPC2 data found that
it was an effective method for removing CRs even on undersampled data
\citep{vandokkum01}.  We therefore adopted the same 'la\_cosmic' IDL
code for our data reduction. 
However, we found two problems when the method was applied to our data. 
First, while the code
appeared to be very effective at removing cosmic rays, there were also
a significant number of obvious false detections in high brightness
regions.  Second,
we found that removed cosmic rays still left small halos behind.  To
address the false detections we set 
very stringent limits for what was counted as a cosmic ray (sigclip=6,
objlim=8).  To remove the residual halos, we expanded the cosmic ray removal
to the four nearest neighbors of a detected cosmic ray. Although this
undoubtedly removed some pixels unaffected by CRs it also eliminated all
signs of the halos around removed CRs.  
In total, the altered 'la\_cosmic' routine resulted in the removal and
interpolation over $\sim$150,000 pixels per science image in the FLT
files.  However it also left behind some obvious cosmic ray features.
We note that when we had only one exposure in a filter, we used less
stringent parameters for selecting cosmic rays, which may have
resulted in the removal of real sources.  However, these images were
not used for source detections so the only affect of this would be
occasionally spurious photometric results.

The remaining cosmic rays were removed in the drizzling process.  We
ran the PyRAF 'multidrizzle' task to create 4350 $\times$ 4350 pixel
images.  Due to our 
crowded fields, the default sky subtraction produced obvious
mismatches in sky brightness between chip quadrants, requiring us to
use skytype='single', which applies a single sky value to
each CCD image.  We also found determining the sky value using the
mode (skystat='mode') rather than using the median (which is the
default), gave better matched sky values.  Despite these measures,
some of our 
most crowded fields (e.g. NGC 4631) show a slight mismatch in sky
brightness between the two CCDs.  This should not affect the
identification or photometry of stars in these images because both use
locally determined sky values.  Comparing the 'median',
'crreject' and 'minmed' modes of image combination, we found the
default 'median' mode surprisingly yielded the best results for cosmic
ray removal.  The 'lanczos3' kernel was used for creating the final
images with a pixel scale of 0.05 arcseconds.  We found that the
'lanczos3' kernel gave a sharper image than the default 'square'
kernel.  Specifically, within a two pixel aperture,
stars in an image made with the 'lanczos3' kernel had $\sim$5\% more
flux than in an image made with a 'square' kernel.  For apertures
larger than four pixels, fluxes in the two images were the same.  
All further analysis is done with the final, drizzled images.  

\subsection{Object Identification \& Aperture Photometry}

All of the galaxies in the sample show a resolved stellar component.
Individual stars were detected using the FIND routine in DAOPHOT II
\citep{stetson87} with a threshold of 5$\sigma$ over the local
background.  The routine initially returned anywhere from 15,000 sources in
IRAS~06070-6147 to 400,000 sources in NGC~55.  The final number of
sources, shown in Table~3, count only sources detected in both
filters.  For the 1$\times$2 fields, 
we detected objects only in the F814W filter and then used that source
list for both filters.  

We performed aperture photometry using the PHOT routine with an
aperture of 4 pixels and a sky annulus between 11 and 14 pixels.
We then compiled the photometry in both filters and required objects
be detected independently in both filters for the 2$\times$2 fields.
This reduced the number of objects by 40-75\% from the original source
lists.  We note that there was a $\sim$0.3 pixel positional offset between the
F606W and F814W drizzled images in all the observations.  After
correcting for this offset we matched objects within 1 pixel of
eachother in the two filters.  In the very small number of cases
where more than one source was found within one pixel, the brighter match
was used.  As noted above, in the 1$\times$2 fields, sources were
found only in the F814W band -- the positions from the F814W image were
then used for both filters to determine the aperture magnitudes.

We transformed all of our photometric data to the VEGAmag system using
provided zeropoints (ACS ISR 2004-8, De Marchi et al.) and aperture
corrections (M. Sirianni, private communication).  The zeropoints provided are for an infinite
aperture and the aperture corrections give the aperture correction
from a 10 pixel to infinite aperture.  We derived a correction from a
4 to 10 pixel aperture using isolated bright stars from the 2$\times$2
fields.  The
resulting corrections showed very little ($<$0.01 mag.) scatter and
are shown in 
Table~4 along with the zeropoints and 10$-\infty$ aperture
corrections.

\begin{deluxetable*}{lcccc}
\tabletypesize{\scriptsize}
\tablecaption{Photometric Constants}
\tablehead{
     \colhead{Filter}  &
     \colhead{Zeropoint} &
     \colhead{10$-\infty$ Ap. Cor.}  &
     \colhead{4$-$10 Ap. Cor.} &
     \colhead{PSF-10 Cor.} \\
     \colhead{}  &
     \colhead{[mag]} &
     \colhead{[mag]}  &
     \colhead{[mag]} &
     \colhead{[mag]} 
}
\startdata
F606W & 26.398 & 0.091$\pm$0.001 & 0.079$\pm$0.006 & 0.017$\pm$0.027 \\
F814W & 25.501 & 0.090$\pm$0.001 & 0.095$\pm$0.009 & 0.019$\pm$0.025 \\
\enddata
\end{deluxetable*}

\subsection{PSF photometry}

All of the galaxy images are very crowded, greatly reducing
the accuracy of aperture photometry.  To improve the
photometry, we modeled the PSF in our drizzled
fields.  An adequate model for the ACS/WFC PSF must allow for spatial
variations of the PSF over the wide field of view.  Fortunately,
DAOPHOT II has the capability to model the PSF as a combination of an
analytical function and empirical sub-pixel corrections that depend
quadratically on the position in the field.  

Initially we attempted to model the PSF in each field independently, but
found that the derived PSFs were greatly limited by the small numbers
of bright, isolated stars (typically 50) available in the large
4350$\times$4350 pixel frames.  Instead we assumed that the PSF was roughly
constant over the course of our observations and
constructed a patchwork image in each filter using bright, isolated
stars from the 
2$\times$2 fields.  The resulting images contained
$\sim$700 stars and 
enabled more accurate determination of the spatial variation of the PSF.  
Out of the options given by DAOPHOT II, the best fitting analytical
functions in both filters were Moffat functions, ${\rm f(x) =
  1/(1+(r/\alpha^{2}))^{\beta}}$.  In the F606W
filter, $\alpha = 0.76$ and $\beta = 2.5$, while in F814W
$\alpha = 0.63$ and $\beta =  1.5$.  The FWHM of these analytical
functions is 0.86 pixels (0.043 arcsec) in F606W and 0.97 pixels (0.049
arcsec) in F814W.  Based on a comparison between a variable and
constant PSF, the spatial corrections in our PSF affect the photometry by at
most 0.03 magnitudes.  The correction of PSF magnitudes to 10 pixel
aperture photometry is given in Table~4.  The accuracy of these
corrections is $\sim$0.03 magnitudes.  We treat this as a systematic
error, and do not include it in individual errors of the stars.  The computed
PSF has a radius of 7 pixels (0.35 arcsec).  

Figure~1a shows the resulting PSF magnitude color-magnitude diagram
(CMD) for the NGC~55-DISK field.  The NGC~55-DISK field CMD is the
cleanest of our CMDs due to the galaxy's proximity and the field's
relative lack of dust.  The clear features seen in this CMD, which
will be discussed in \S3, demonstrate the quality of our photometry.

\begin{figure*}
\epsscale{1.0}
\plotone{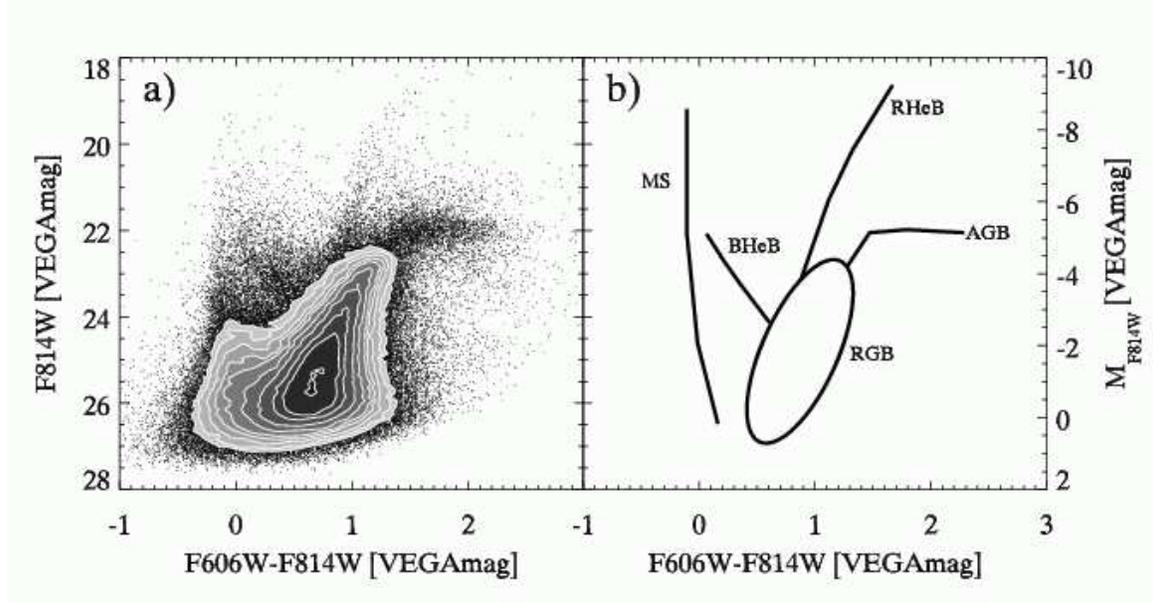}
\caption{{\it (a)} Color-Magnitude Diagram (CMD) of the 253,108 stars in the 
  NGC 55-DISK field.  Contours
  are used where the density of points becomes 
  high.  The contours are drawn at densities of 75, 100, 150, 200,
  250, 350, 500, 750, 1000, and 1500 
  stars per 0.1 magnitude and color bin.  {\it (b)} Cartoon CMD
  showing the location of Main Sequence 
  (MS), Red Giant Branch (RGB), Asymptotic Giant Branch (AGB), Red
  Helium Burning Sequence (RHeB), and Blue Helium Burning Sequence
  (BHeB) each of which is described in \S3.  }
\end{figure*}

\subsection{Artificial Star Tests}

To assess the completeness limit as a function of position,
magnitude, and color, we inserted large numbers
of artificial stars in our images and then attempted to recover them
with our data reduction pipeline.  These stars were input in grids
separated by 20 pixels with an overall random position offset.
In this way, $\sim$38,000 stars were added to a field, each with a
random F606W VEGAmag between 18 and 29 and a F606W-F814W color between
-1.0 and 3.0.  We then ran the stars through a pipeline identical to
the one outlined above for finding stars and determining their
photometry.  This was done with different random position offsets to generate
as many as four million artificial stars distributed throughout each
field.  The large
computing time required to run these tests was enabled through use of
the distributed computing system, Condor\footnote{The
  Condor Software Program (Condor) was developed by the Condor Team at
  the Computer Sciences Department of the University of
  Wisconsin-Madison.  All rights, title, and interest in Condor are
  owned by the Condor Team.}.

\begin{figure}
\plotone{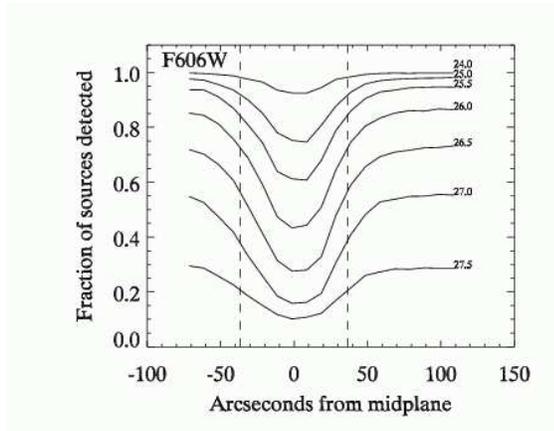}
\caption{The fraction of artificial stars detected as a
function height above the midplane of the galaxy in NGC 4244.  The
lines give the completeness in 0.5 mag wide bins around a magnitude
shown on the right-hand side.  Dashed lines show $\pm$3z$_{1/2}$, the
minimum height used for the TRGB determination in \S4.} 
\end{figure}

Figure~2 shows the completeness as a function of scale height in
NGC~4244 for stars at a variety of magnitudes.  Away from the plane of
the galaxy, in a non-crowded area, the 50\% completeness level of our
data is $\sim$27 mag in F606W and $\sim$26 mag in F814W.  However,
the crowding in the midplane of the galaxy drastically reduces the
completeness, by as much as 35\% at moderate magnitudes.  Note that
Figure~2 contains the results from artificial stars of all colors,
but for a specific range in color the completness can be higher or lower.
For instance, the F814W 50\% completeness in non-crowded areas at
F606W-F814W between -1 and 
1 is $\sim$26.5, but rises steeply at redder colors.  Figure~3 shows a
typical 20\% completeness line as a function of color in each of the
color-magnitude diagrams. The scatter from galaxy to galaxy in the
completeness values is a few tenths of a magnitude.  

\subsection{Stellar Models \& Photometric Transformations}

Critical to the interpretation of our data is the ability to connect the ACS
VEGA magnitude system for the F606W and F814W filters
with existing stellar models and previous data.  We rely on two
sources for performing this photometric transformation.  First, we used a set of the Padova isochrones and
single stellar population models 
transformed into the ACS VEGAmag system using
synthetic spectra and ACS/WFC instrument response
(L. Girardi, private communication).  These 
models are the standard \citet{girardi00} isochrones with tracks at
Z=0.0001, 0.0004, 0.001, 0.004, 0.008, \& 0.019 ([Fe/H]=-2.3, -1.7,
-1.3, -0.7, -0.4, 0.0) with ages between 4 Myr and 18 Gyr.  

Second, in very limited cases, we used transformations from the ACS
VEGAmag system to Johnson V \& I provided by M. Sirianni (private
communication). 
However, these transformations are potentially inaccurate as they
attempt to infer
the spectrum of a star from limited color information and are valid
only for -0.5$<$V-I$<$2.3.  

The VEGAmag F606W-F814W color is similar to Johnson V-I, but has a 
somewhat smaller wavelength span, meaning that the differences in
color are somewhat more moderate.  For instance, Red Giant Branch
stars are $\sim$0.3 magnitudes bluer in F606W-F814W [VEGAmag] than in
V-I.  Unless explicitly stated otherwise, all magnitudes in the paper
are in the VEGAmag system.

\section{Color-magnitude Diagrams}

In the sample as a whole we have determined photometry for over 1.2
million objects (see Table~3), a vast majority of which are stars in
the sample galaxies.  

\begin{figure*}
\plotone{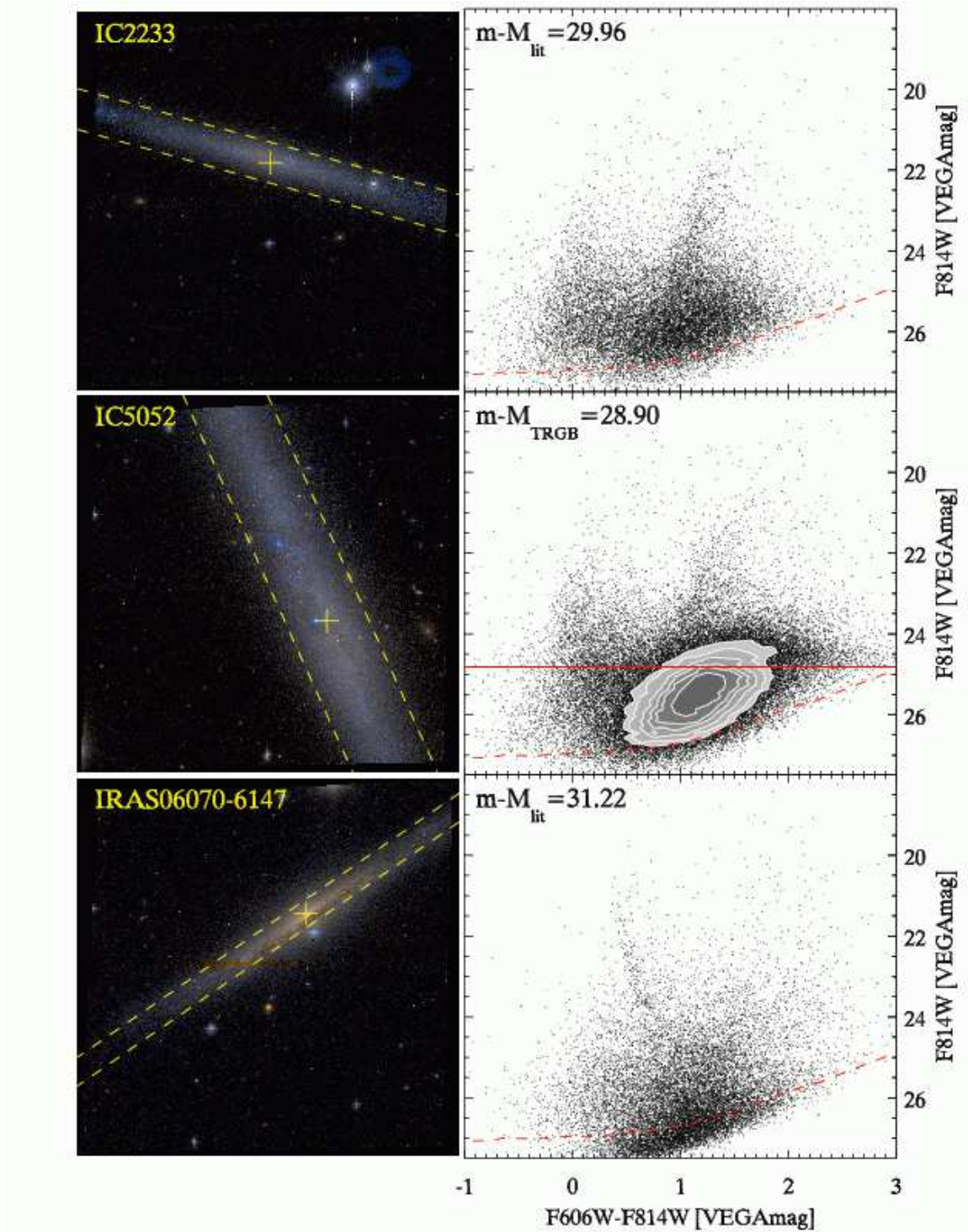}
\caption{{\it (left)} -- Combined F606W and F814W
  images of each field. 
  Dashed lines indicate 3z$_{1/2}$ and cross indicates galaxy center
  as determined from fits to K band 2MASS data (see Appendix).  The images
  are 218\arcsec$\times$218\arcsec.  The box shown in the
  NGC~4631-DISK field gives the location of image shown in Fig.~4.
  Note that color images of these pictures are presented in the
  electronic version.  
  {\it (right)} --
  Color magnitude diagrams of all the objects in each field.  Contours
  are shown where densities exceed 75 stars per 0.1 magnitude and
  color bin.  Contours have the same levels as those shown in Fig.~1.
  The dashed line indicates the typical 20\% completeness limit.  The
  solid horizontal line (when present) indicates the TRGB.  The nearly
  vertical feature at F606W-F814W$\sim$0.4 in IRAS 06070-6147 is due
  to LMC field stars.  Distance moduli m-M$_{\rm lit}$ or
  m-M$_{\rm TRGB}$ come from Tables~1 and 5.}  
\end{figure*}

\addtocounter{figure}{-1}
\begin{figure*}
\plotone{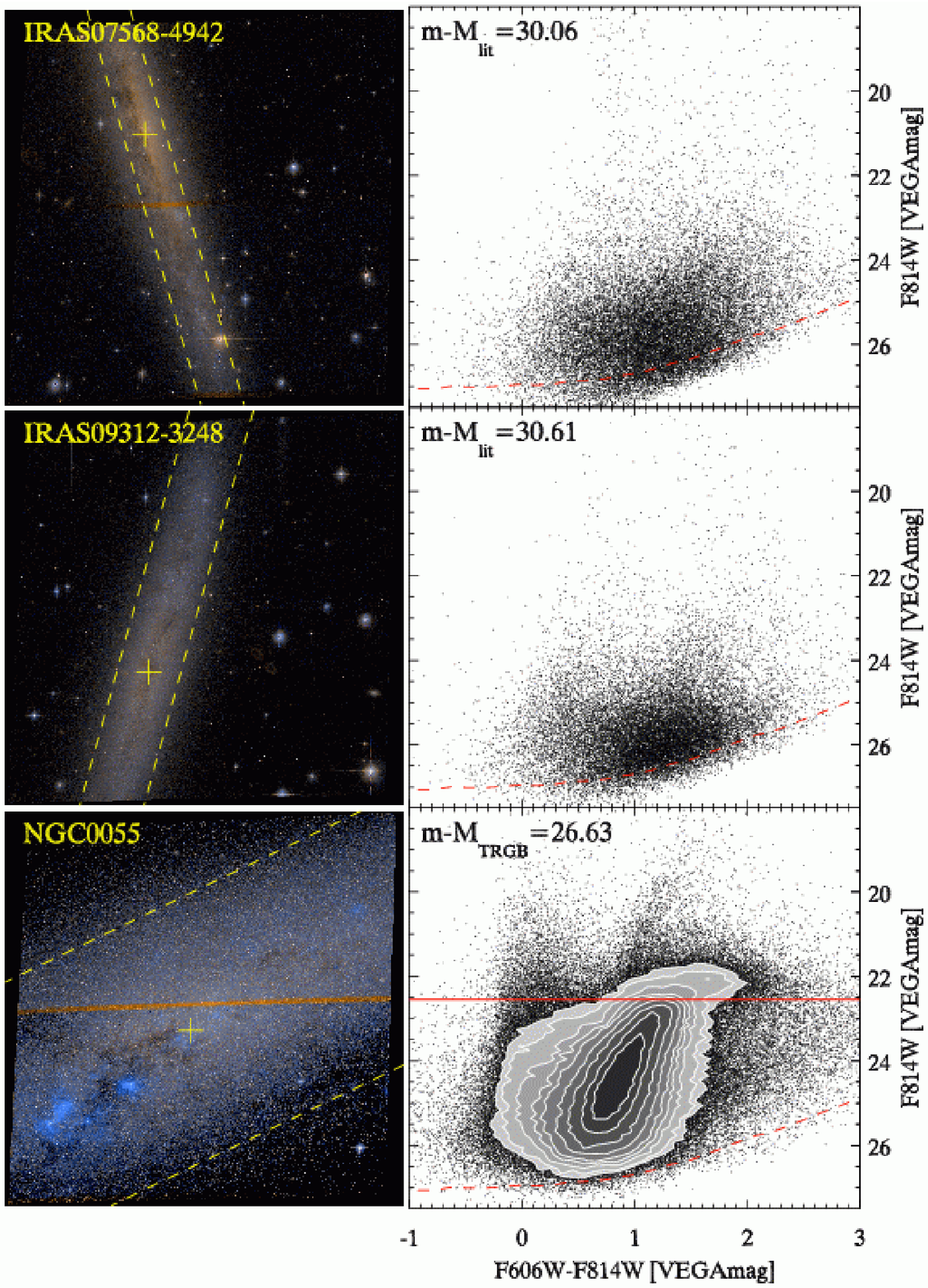}
\caption{{\it continued}}
\end{figure*}

\addtocounter{figure}{-1}
\begin{figure*}
\plotone{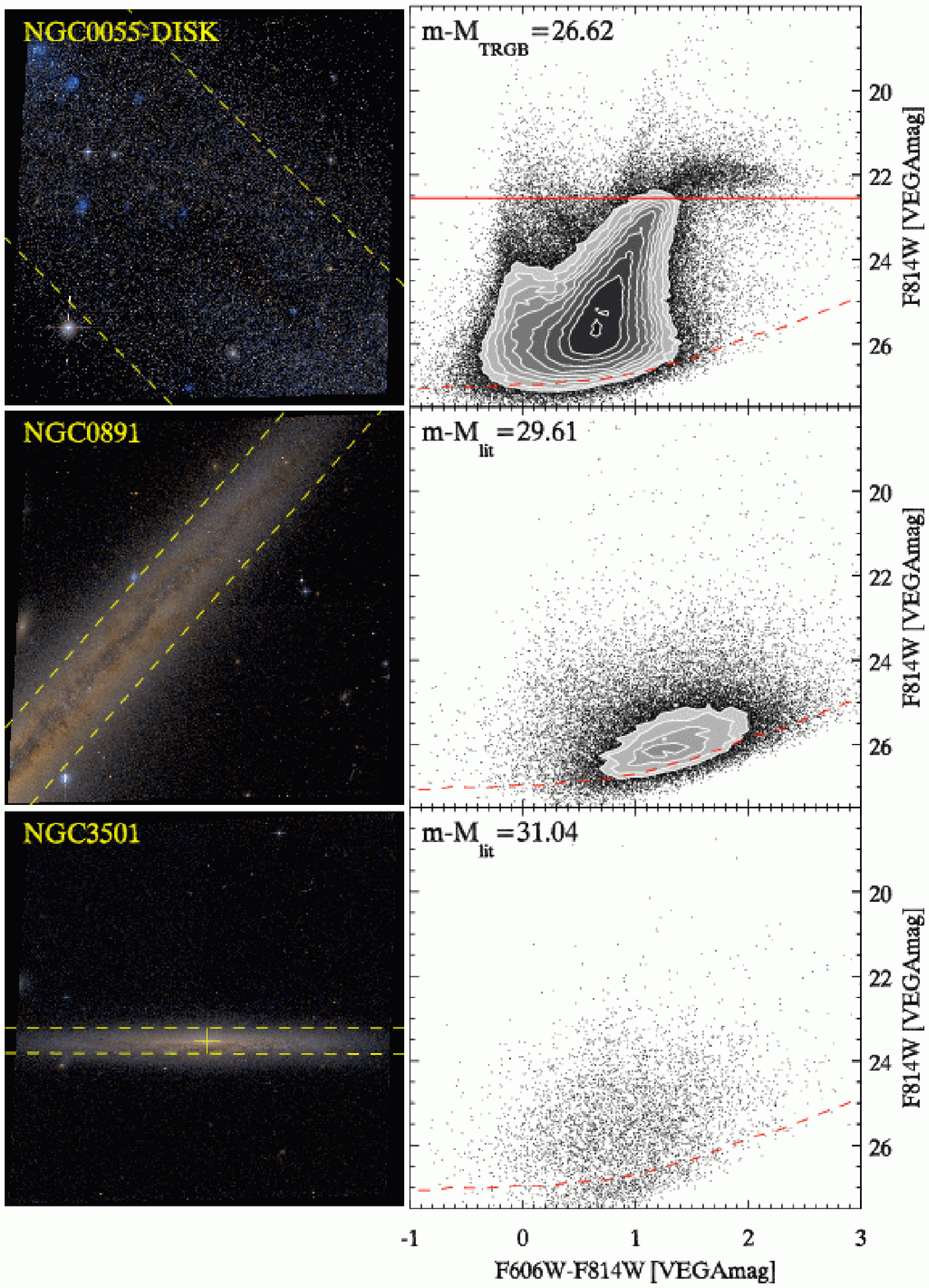}
\caption{{\it continued}}
\end{figure*}

\addtocounter{figure}{-1}
\begin{figure*}
\plotone{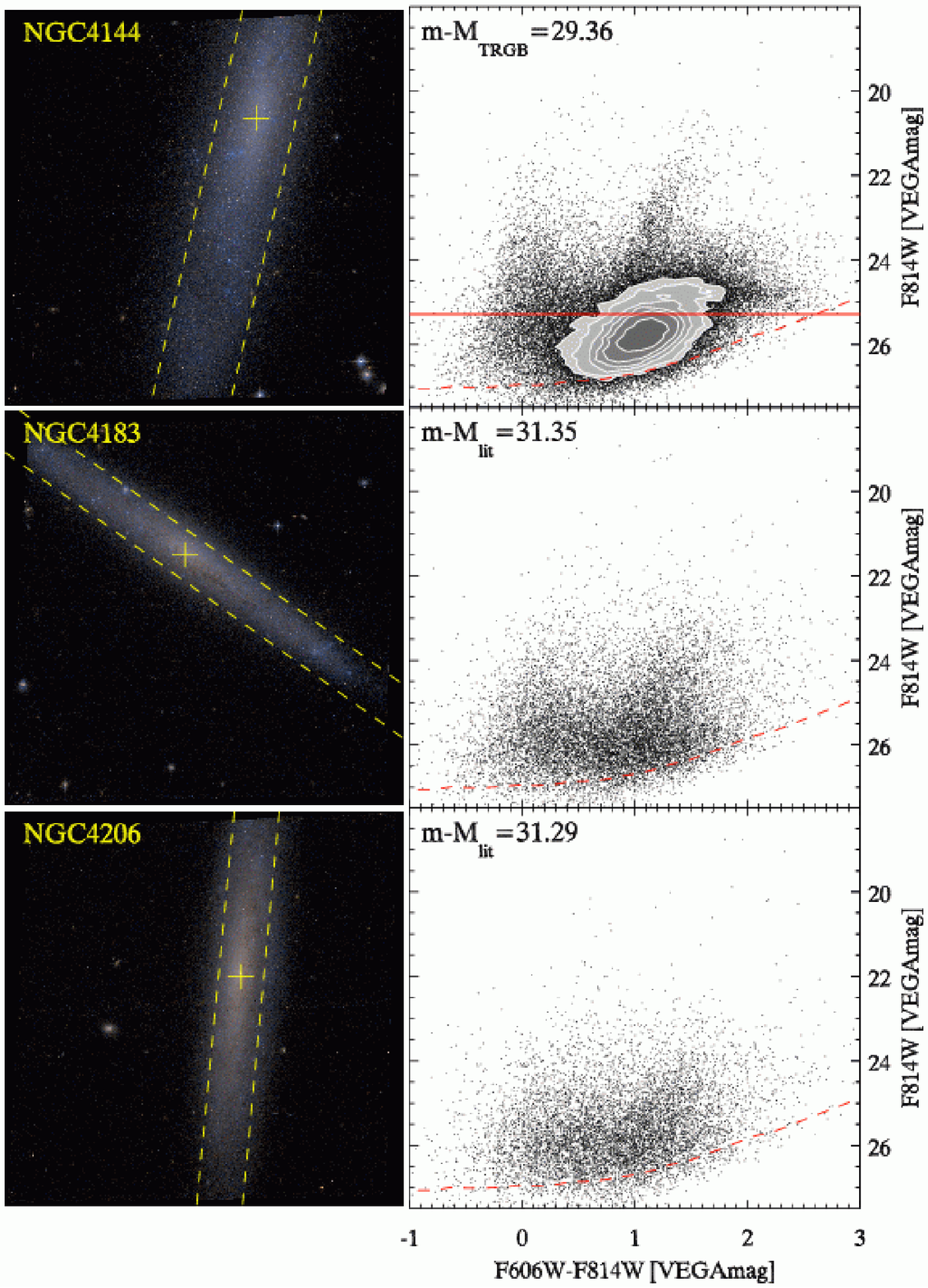}
\caption{{\it continued}}
\end{figure*}

\addtocounter{figure}{-1}
\begin{figure*}
\plotone{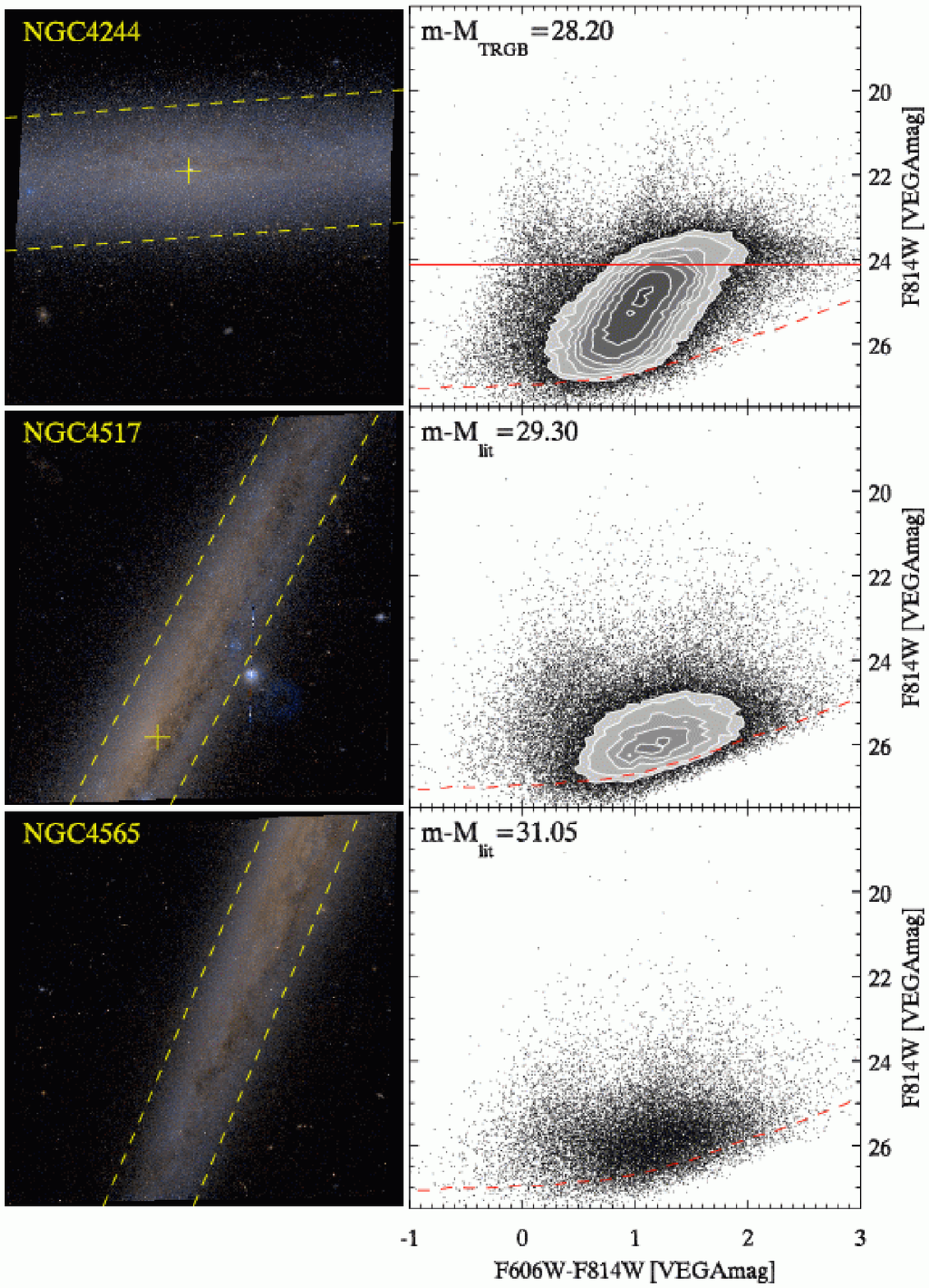}
\caption{{\it continued}}
\end{figure*}

\addtocounter{figure}{-1}
\begin{figure*}
\plotone{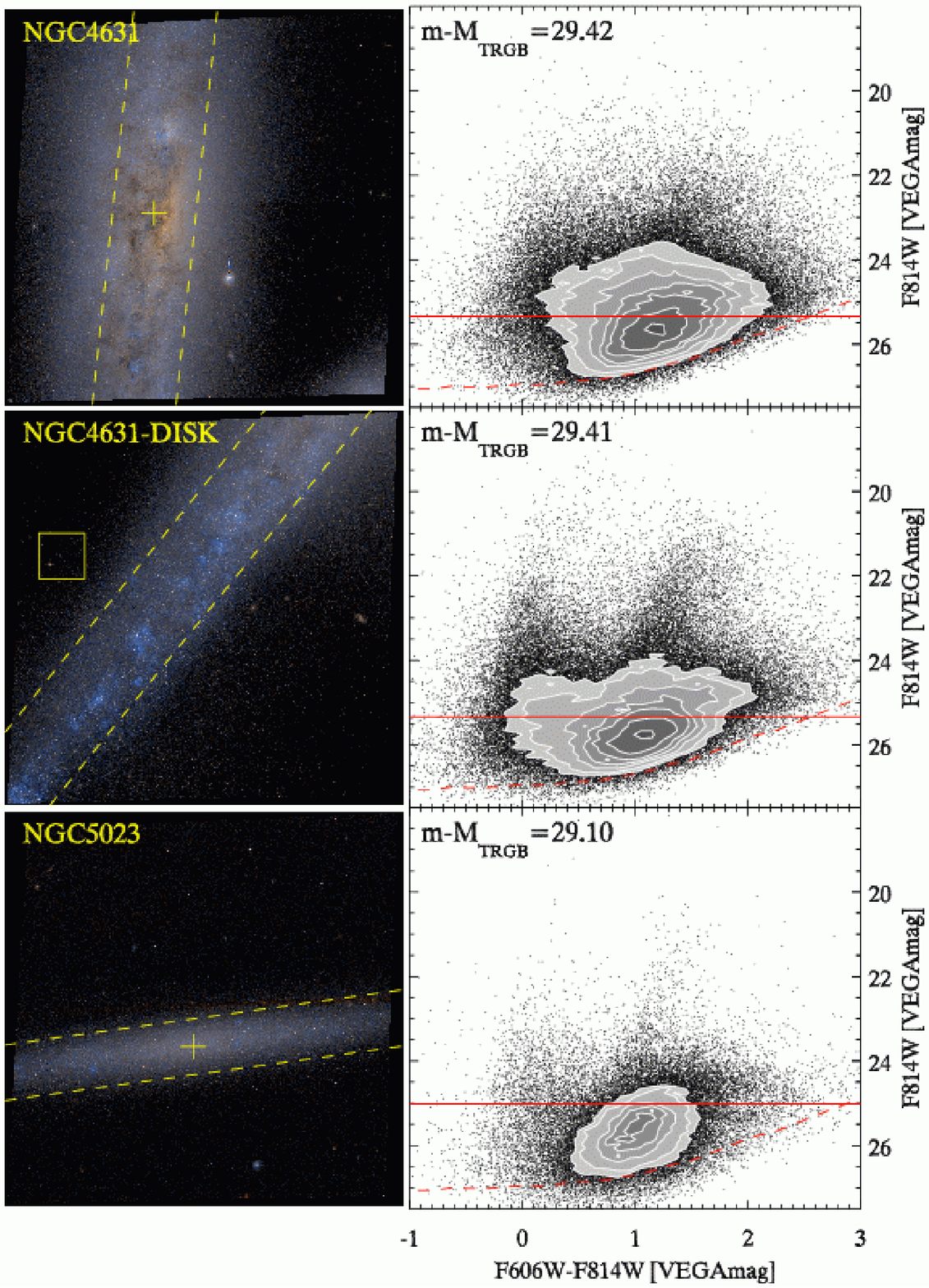}
\caption{{\it continued}}
\end{figure*}

Figure~3 shows images and color-magnitude diagrams (CMDs) of each of
the galaxies.  The CMDs show all the stars detected in each field.
Of all the CMDs in our sample, the NGC~55-DISK CMD reaches the
faintest absolute magnitudes.  In Figure~1b we illustrate the
distinct features visible in its CMD.  Many of the other CMDs in
Figure~3 share these features at fainter apparent magnitudes.  These
visible features are due to the following stellar populations: 
\begin{itemize}
\item Main Sequence (MS) - At the bright absolute magnitudes at which
  we are observing (M$_{\rm F814}<0$), MS stars are younger than a
  few 100 Myr.  At the bright end, the MS and Blue Helium Burning
  sequence overlap.  
\item Red and Blue Helium Burning Sequences (RHeB and BHeB) -  These
  sequences contain young massive stars undergoing core helium
  burning.  Along both these sequences, the ages and masses of the 
  stars are proportional to the luminosity, with younger, more 
  massive stars populating the bright end.  The faintest stars along the RHeB
  have an age of $\sim$200 Myr, while
  the BHeB branch extends down to $\sim$600 Myr
  \citep{dohm-palmer02a,dohm-palmer02b}.  
\item Asymptotic Giant Branch (AGB) - The AGB contains stars of
  intermediate age, after completion of their core-helium burning
  stage.  The 
  observed portion
  of the AGB is brighter and redder than the RGB.  Based
  on the Girardi isochrones, the AGB stars observed here mostly have
  metallicities between [Fe/H]=-0.7 and 0.0 and ages between 100 
  Myr and a few Gyr.  At absolute magnitudes fainter than the red giant 
  branch tip M$_{\rm F814W}\sim-4$ the AGB stars cannot be
  distinguished from the RGB stars
\item Red Giant Branch (RGB) - The RGB is
  dominated by old stars undergoing shell-hydrogen burning.  These
  stars have ages $>1$ Gyr and represent the oldest of the components
  visible in our CMDs.  The exact position of the RGB depends
  primarily on metallicity, with more metal-rich stars being redder
  and fainter in our CMDs.  
\end{itemize}

Inspection of Figure~3 shows that most of these components
are clearly visible in IC 2233, IC 5052, NGC 55 (both fields), NGC
4144, NGC 4244, NGC 4631-DISK, and NGC 5023.   These CMDs clearly show
evidence for both young and old stellar populations.  The effects of
dust are also obvious when comparing the color magnitude diagrams of
the central NGC 55 and NGC 4631 fields with those taken further out in
the disk (NGC 55-DISK and NGC 4631-DISK).  In both cases the central
CMDs show significant redward blurring of the MS and RHeB, both of
which are more clearly detected in the less obscured '-DISK' fields.  

We note that a vast majority of 
stars detected in each field appear to be in the target galaxies.
This can be seen in the clarity of their CMDs and from the spatial
distribution of the stars (see Paper II).  A significant population of
Galactic foreground
stars are visible in the CMDs of the three galaxies with the lowest galactic
latitudes (see Table~1), IRAS 07568-4942, IRAS 09312-3248 and NGC
891.  They are seen most prominently at F606W-F814W $>$ 0.5 and F814W $<$
23. 

Before discussing distance measurements using the RGB tip in
\S4, we briefly discuss two unusual features seen in the CMDs of IRAS
06070-6147 and NGC 4631-DISK.  
All further interpretation of the CMDs will be presented in Paper II,
in which we examine the vertical distribution of the different
stellar populations and find evidence for old and metal poor
thick-disks in these galaxies.

\subsection{A Detection of the Large Magellanic Cloud (LMC)}

One unusual feature is found in the CMD of IRAS 06070-6147.  In the
CMD, a clear locus of points can be seen extending between 20th and
24th magnitude and at a color of $\sim$0.5.  These are field
LMC stars found $\sim$9 degrees (8 kpc) from the LMC center, near
the outermost clusters associated with the LMC \citep{irwin91}.

These stars fit well to isochrones with a distance modulus of 18.5
\citep{alves04}, 
an age of a few Gyr, and metallicities of [Fe/H]=-0.4 to 0.0.  This
provides a serendiptous confirmation of the accuracy of our
photometric calibration and isochrone bandpass transformation.  It
also agrees with recent work by \citet{gallart04} which suggests that
the extended 'halo' of the LMC is made up of stars as young as $\sim$2.5
Gyr.  

\subsection{A Young Dwarf Galaxy Candidate in NGC 4631}

Inspection of the NGC 4631-DISK field image showed a small object
marked by a grouping of blue stars sitting significantly off 
the plane of the disk.  This object is shown in Figure~4 and
is centered at $\alpha=$12h41m50.30s and  $\delta=$32$^\circ$31'1.9\arcsec.
The object seems to be contained within a radius of
4.5\arcsec as shown in Fig.~4.  Assuming a distance modulus of
29.42 to the object, this gives a radius of $\sim$170 pc.  The object is
located $\sim$2 kpc above the plane of NGC 4631.  Based on its size,
which is too large to be a star cluster, we identify this object
as a dwarf galaxy candidate, NGC~4631~DwA (dwarf A).  

\begin{figure}
\plotone{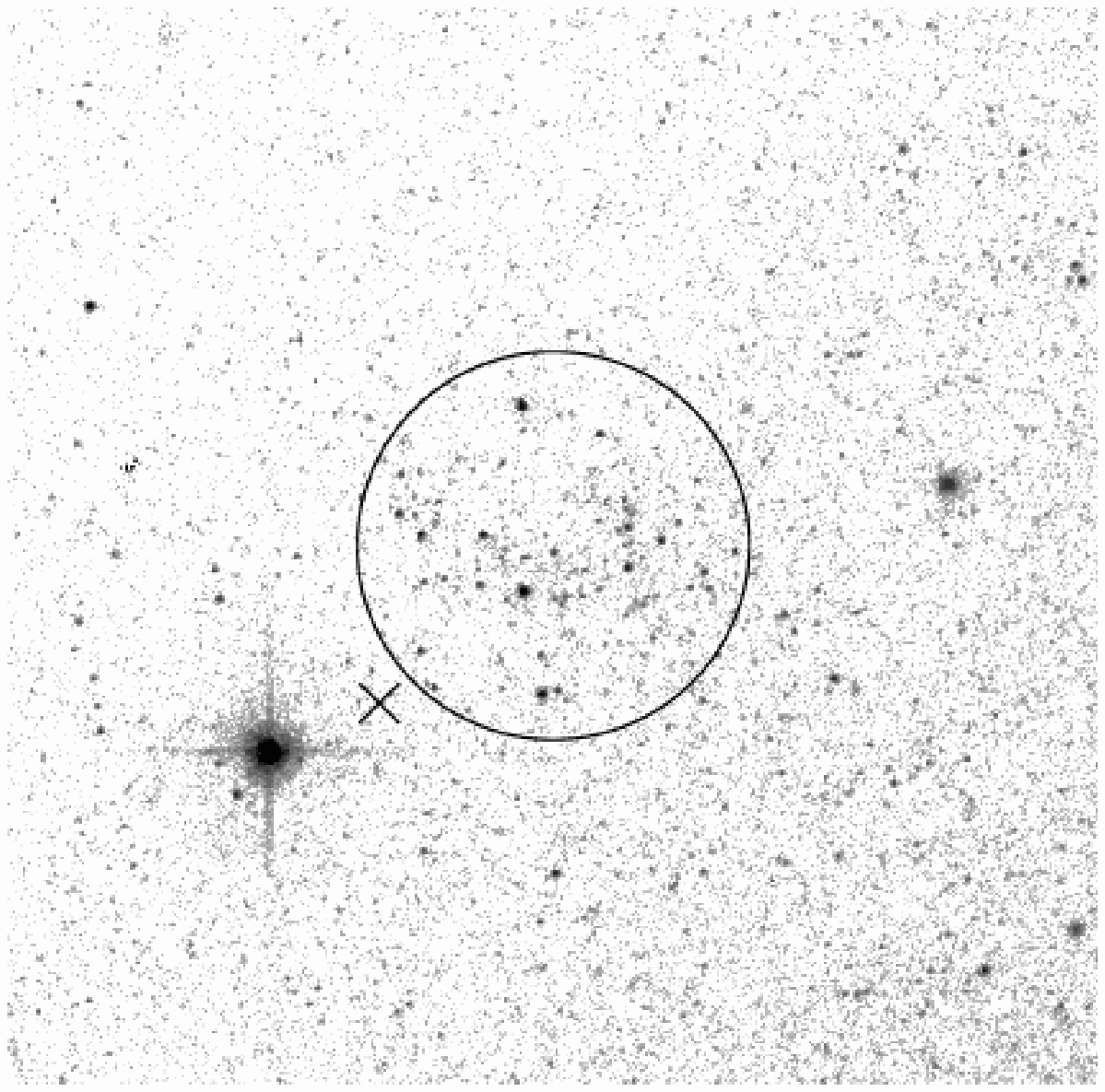}
\caption{An F606W image showing a close-up of the section of NGC 4631-DISK
  shown as a box in Figure~3.  The circle has a radius of 4.5\arcsec
  and is centered at $\alpha=$ 12h41m50.30s and
  $\delta=$32$^\circ$31'1.9\arcsec.  The X marks the location of
  previously identified galaxy NGP9 F268-1993301 \citep{odewahn95}.
  A clear overdensity of stars is seen inside the circle.}
\end{figure}

\begin{figure}
\plotone{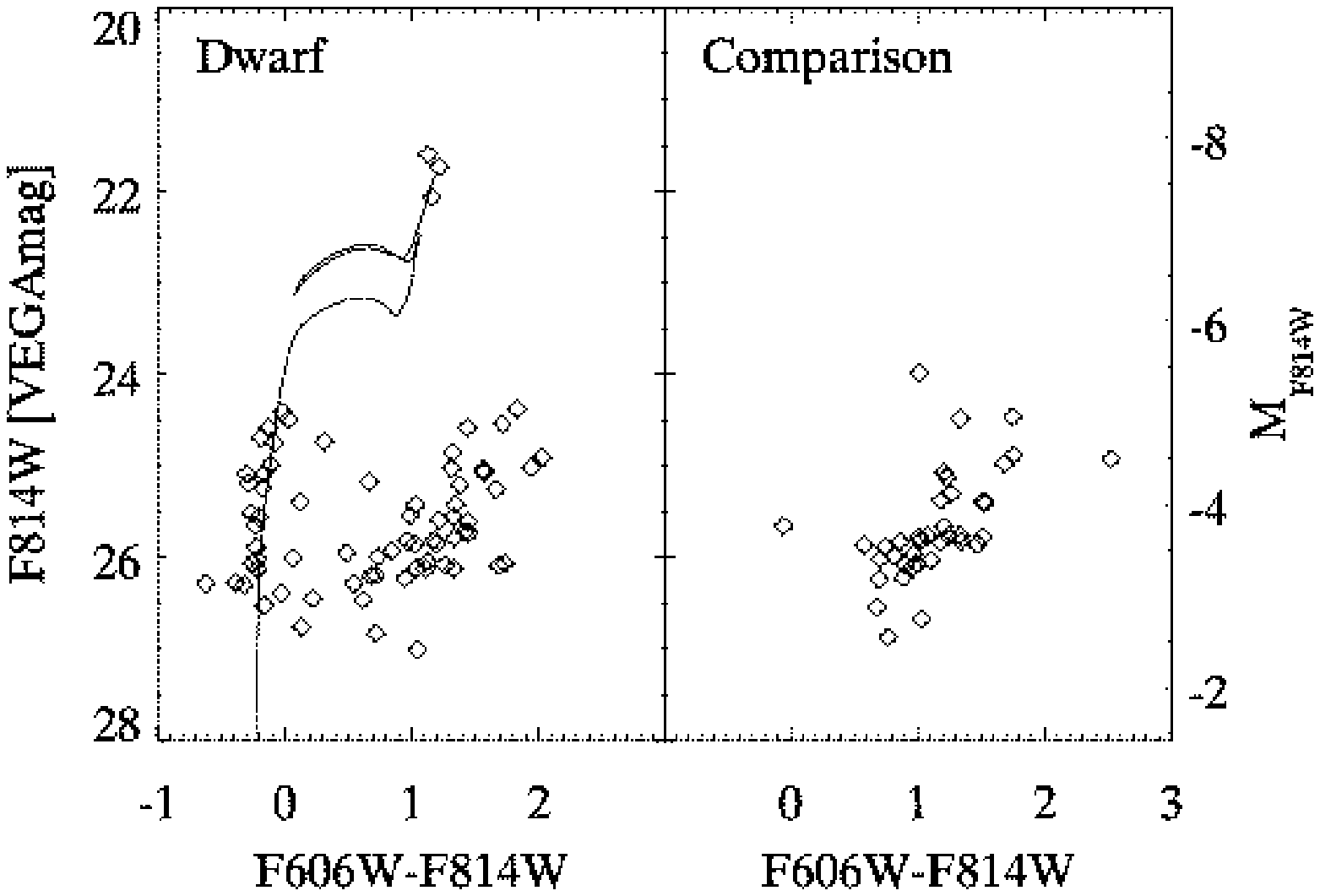}
\caption{{\it (left)} -- A CMD of the stars inside
  the circle shown in Figure~4.  Overplotted is an isochrone with an
  age of 30 Myr and [Fe/H]=-0.7.  {\it (right)} -- The CMD of a comparison
  field located at the same galactic latitude as the dwarf, but offset
  a short distance in the longitudinal direction.  An absolute
  magnitude scale is shown assuming m-M=29.42.}
\end{figure}

This dwarf may have
been previously discovered by \citet{odewahn95} with the name
NGP9~F268-1993301 using POSS-I plates.  However, their position 
($\alpha$=12h41m49.89s $\delta$-+32$^{\circ}$31'00.5\arcsec, shown with
an X in Fig.~4) and red
color (O-E$\sim$2) suggests that they suffered from confusion with the
bright red star (most likely a foreground object) seen to the lower left
of the circle in Figure~4.    

The left panel of Figure~5 shows the color-magnitude diagram of the
stars in the circle shown in Figure~4.  For comparison, we show a
field CMD at the 
same height above the plane of NGC 4631 but offset in the longitudinal
direction.   
NGC~4631~DwA shows a clear overdensity of blue stars.  To show this
quantitatively, we compare the stars in the dwarf to those within
the same range of height above the plane, but lying outside the
dwarf, yielding an area
$\sim$18 times the area of the dwarf.  We find 26 stars bluer
than F606W-F814W=0.3 associated with the dwarf, but only 15 in the
entire area outside the dwarf.  This gives an expected rate $\sim$0.8
stars bluer than F606W-F814W=0.3 in the field of the dwarf, implying
that the dwarf has a $>$30 times 
overdensity of blue stars.  By contrast, the expected rate for
stars redder than F606W-F814W of 0.3 in the field of the dwarf is
$\sim$50 stars, while the dwarf contains only 47.  This suggests that
if any old population exists, it is small enough to contain few stars
near the tip of the RGB.  The good fit of the
isochrone shown in Figure~5 to the blue stars and the bright red stars
(probably RHeB stars) in the dwarf suggests an age of $\sim$30 Myr.
We found that isochrones with a wide range of metallicities and ages
between 10-50 Myr can also fit the data quite well.  These tests also
suggest that the dwarf is located at the same distance modulus
(assumed to be 29.42) as NGC 4631.  

We can also calculate the integrated magnitudes of NGC~4631~DwA to
compare to single stellar population models.  The total flux from the
dwarf is measured by aperture photometry using an aperture of
4.5$\arcsec$ and a sky annulus between 5$\arcsec$ and 6\arcsec.  The contribution
from NGC 4631's field stars is then determined by finding the flux using
the same aperture and sky annulus at 100 random positions
with the same height above the plane, but avoiding
NGC~4631~DwA and the bright star visible just to it's
lower left in Figure~4.  The mean of these we use as a
background and the standard deviation is the error in this background.
The final flux for the dwarf is determined by subtracting off the
background and the final error combines the dwarf's error in the flux
with the error in the background.
Using this error and the background
subtracted flux we find that the F606W magnitude is 20.7$\pm$0.2 (a
5$\sigma$ detection) while the F814W
magnitude is 20.3$\pm$0.4 (a 3$\sigma$ detection).
These magnitudes correspond to surface brightnesses
of 25.2 mag arcsec$^{-2}$ in F606W and 24.8 mag arcsec$^{-2}$ in
F814W.  Assuming an age of 30 Myr and [Fe/H]=-0.7, we find a mass of
2.2$\pm$0.4$\times$10$^{4}$ M$_\odot$ from single the stellar
population models (see \S2.5).  We note that despite being rather
uncertain, the F606W-F814W color of 0.4 matches the expected value from
this single stellar population, suggesting that any existing older 
population does not make a significant contribution to
the total luminosity of NGC~4631~DwA.  The dwarf's absolute 
magnitude, M$_{\rm F606W}=-8.7$ is only slightly brighter than the
recently discovered 'faintest galaxy' Andromeda IX \citep{zucker04}. 

The properties of NGC~4631~DwA -- a radius of 170 pc, an age of $\sim$30
Myr, and a mass of $\sim$2$\times$10$^4$ M$_\odot$ are most similar to an OB
association.  However its location $\sim$2 kpc above the disk makes it
unusual.  If the object is indeed bound it would have a circular
velocity of $\sim$7 km/sec.  However given that the only stars we see
are young, it is possible that it is an unbound and gradually
dissolving object.

One possible explanation for the creation of this object is that it is
caused by a condensation of gas interacting with NGC 4631.  Figures 4
and 5 of \citet{rand94} show five HI spurs surrounding NGC 4631.
Although none of their positions overlap with the position of the
dwarf, two spurs meet up with the disk at a position just opposite the
dwarf.  This explanation for the formation of the object would make it
akin to young tidal dwarf galaxies and clusters observed in the tails of
interacting galaxies
\citep[e.g.][]{schweizer78,duc98,temporin03}. Higher resolution HI
observations would make it possible to determine if the object is
affiliated with a tidal tail.

\section{TRGB distances}

The tip of the red giant branch (TRGB) method can be used to determine
distances to objects with an old and metal poor ([Fe/H] $<$
-0.7) resolved stellar population.  Of the sample of galaxies
presented here, eight fields in six galaxies were close enough
to clearly distinguish an RGB tip.  To determine the TRGB magnitude we
use the discontinuity detection method as presented in \citet{mendez02}.

\subsection{Star selection and method}

Before considering the method by which we determine the TRGB, we need
to refine the sample of stars.  To prevent our result from being
affected by {\it in situ} dust and young stellar populations and to
avoid incompleteness  (Fig.~2), we consider stars only above
3$|$z$_{1/2}|$.  This height is above  
$>$90\% of the K band light on the 2MASS images and is above
any obvious dust features on our images, except in NGC 4631 which is
known to have a large extraplanar dust component \citep{howk99}. 
The CMDs constructed from these stars show the presence of very few young
stars.  This can be seen in Figure~6, which
shows the CMD of stars above 3$|$z$_{1/2}|$ except in NGC 4631 and NGC
4631-DISK fields where we used a  
lower limit of 7$|$z$_{1/2}|$ in order to better avoid the dust,
crowding and young 
stellar populations seen at lower scale heights.  For the NGC 4631
field, we also used an upper limit of 10$|$z$_{1/2}|$ to avoid stars
in the nearby companion, NGC 4627. 

\begin{figure*}
\plotone{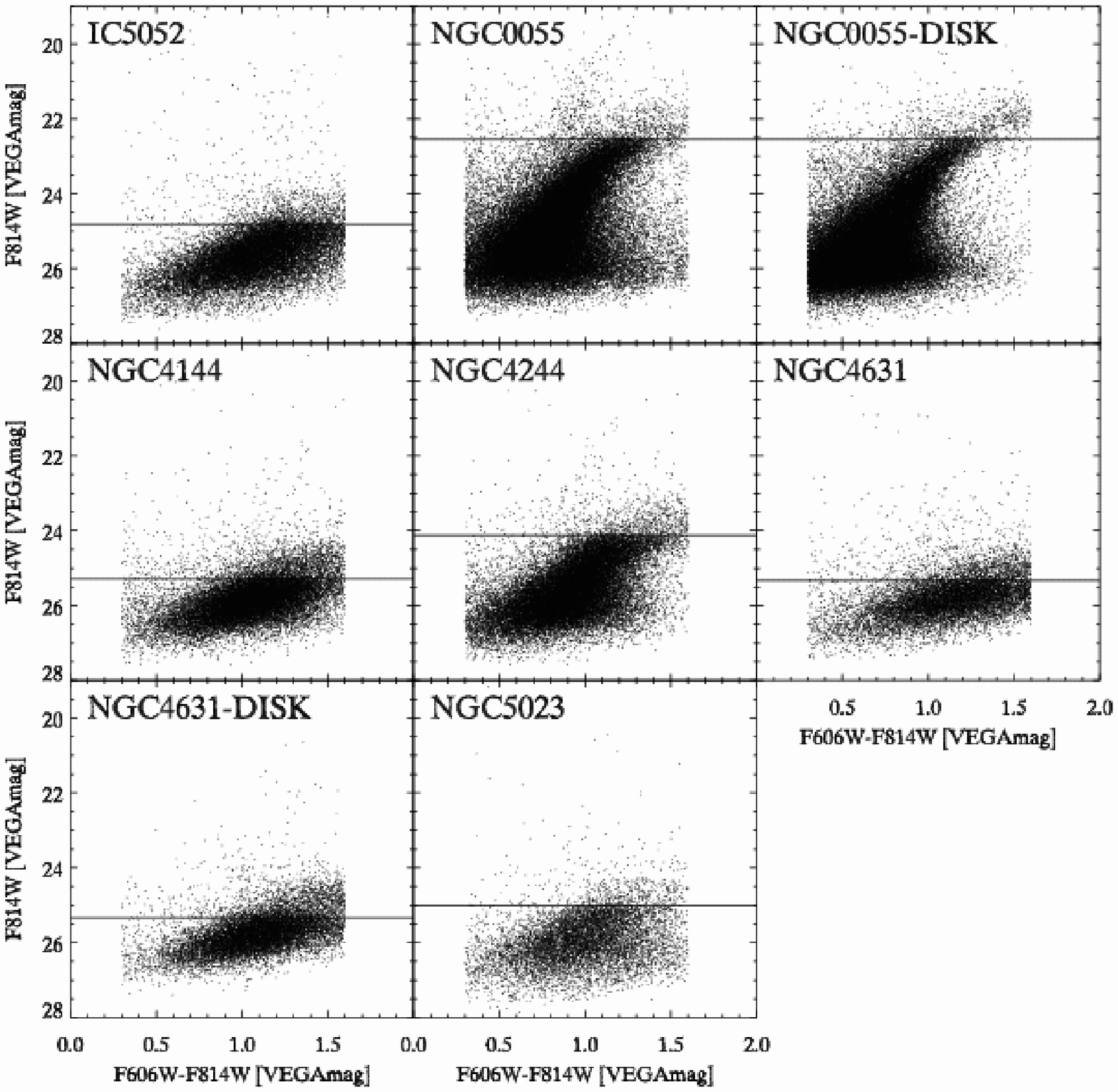}
\caption{The CMDs of stars used in TRGB determination.  The line
  in each pane indicates the magnitude found for the TRGB.  Stars were 
  selected to have 0.3$<$F606W-F814W$<$1.6, and to be well above the
  plane of the galaxy ($>$3z$_{1/2}$ for all galaxies other than NGC
  4631).  These CMDs show a much greater clarity than those in Fig.~3
  because of the reduced effects of dust and crowding.}
\end{figure*}

Comparison of the CMDs in Fig.~6 to the full 
field CMD in Figure~3 shows the lack of young RHeB stars and the greater
clarity of the RGB due to a lack of crowding.  After selecting
these stars, we correct their 
magnitudes for foreground extinction using \citet{schlegel98} values
for E(B-V) and assuming A$_{\rm F814W}$=2.716$\times$E(B-V) and 
A$_{\rm F606W}$=1.796$\times$E(B-V).  Finally, we make a color cut (0.3 $<$
F606W-F814W $<$ 1.6) to eliminate 
contamination of the upper main sequence and redder AGB stars.  

Figure~7 graphically shows the method used to determine TRGB
magnitude and error in each field.  Using the magnitudes and errors
of the selected stars, we construct a 
Gaussian-smoothed luminosity function, to which each star contibutes a
unit area Gaussian with the width determined by the error \citep[see
  Appendix A of][]{sakai96}.  This is shown in the top panels of
Figure~7.  The tip 
of the red giant branch 
can be seen as a sudden increase in the luminosity function at a given
F814W magnitude.  This magnitude can be
quantitatively  
determined using an edge-detection algorithm.  We use the
logarithmic edge-detection algorithm presented in \citet{mendez02}.
The response of this edge-detection algorithm is shown in the middle
panels of Figure~7.  Finally, random errors in the TRGB magnitude are
determined using a Monte Carlo style simulation where the TRGB
magnitude is determined 500 times for stellar magnitudes which have
been randomly resampled with replacement.  We also include the
magnitude errors by adding a Gaussian random error scaled by each
stars photometric error.  The width of a Gaussian fit to the
determined TRGB magnitudes gives us the uncertainty of the derived
TRGB value \citep{mendez02}.  A histogram of these Monte Carlo tests
can be seen in the bottom panels of Figure~7. 

\begin{figure*}
\plottwo{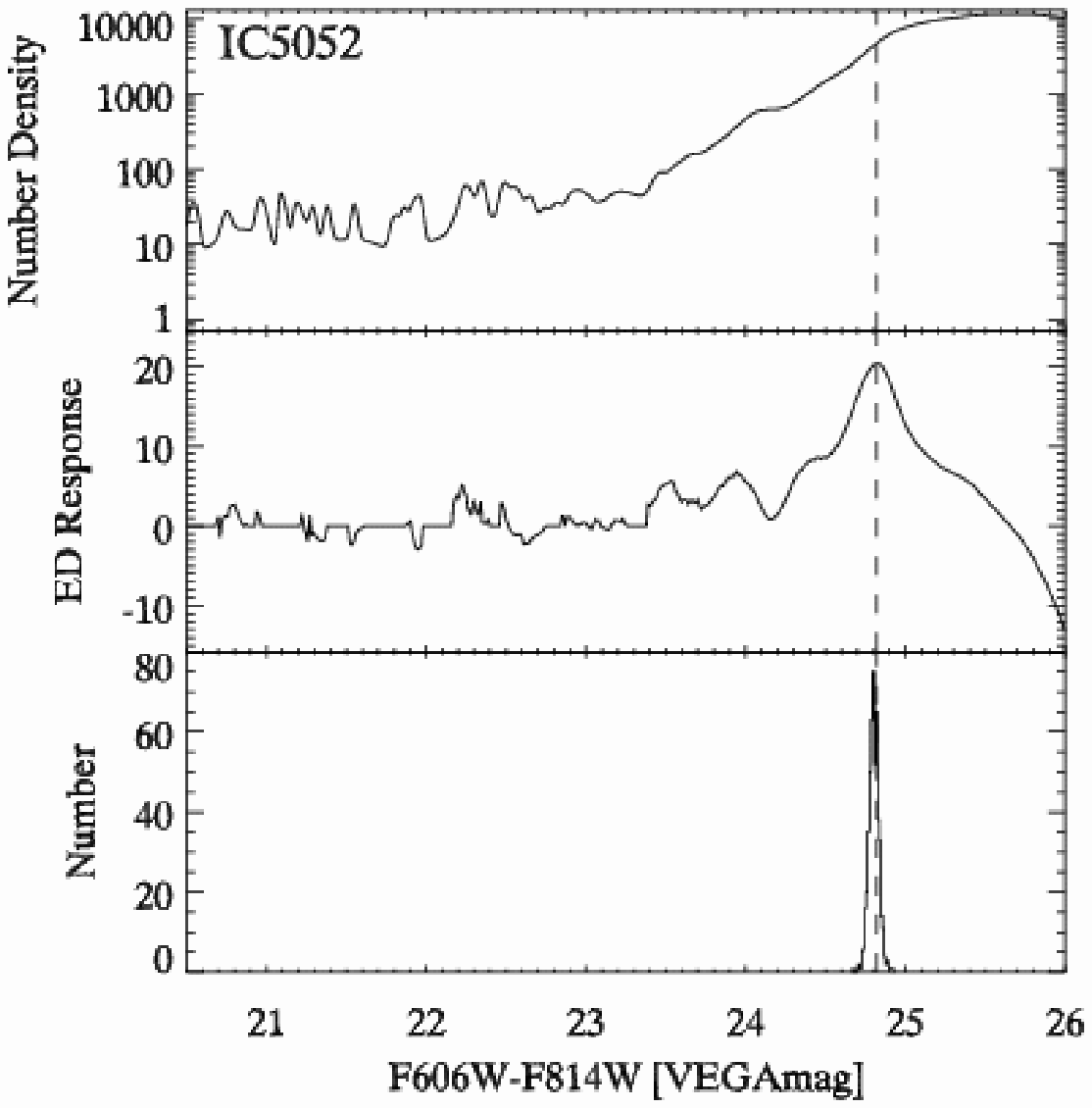}{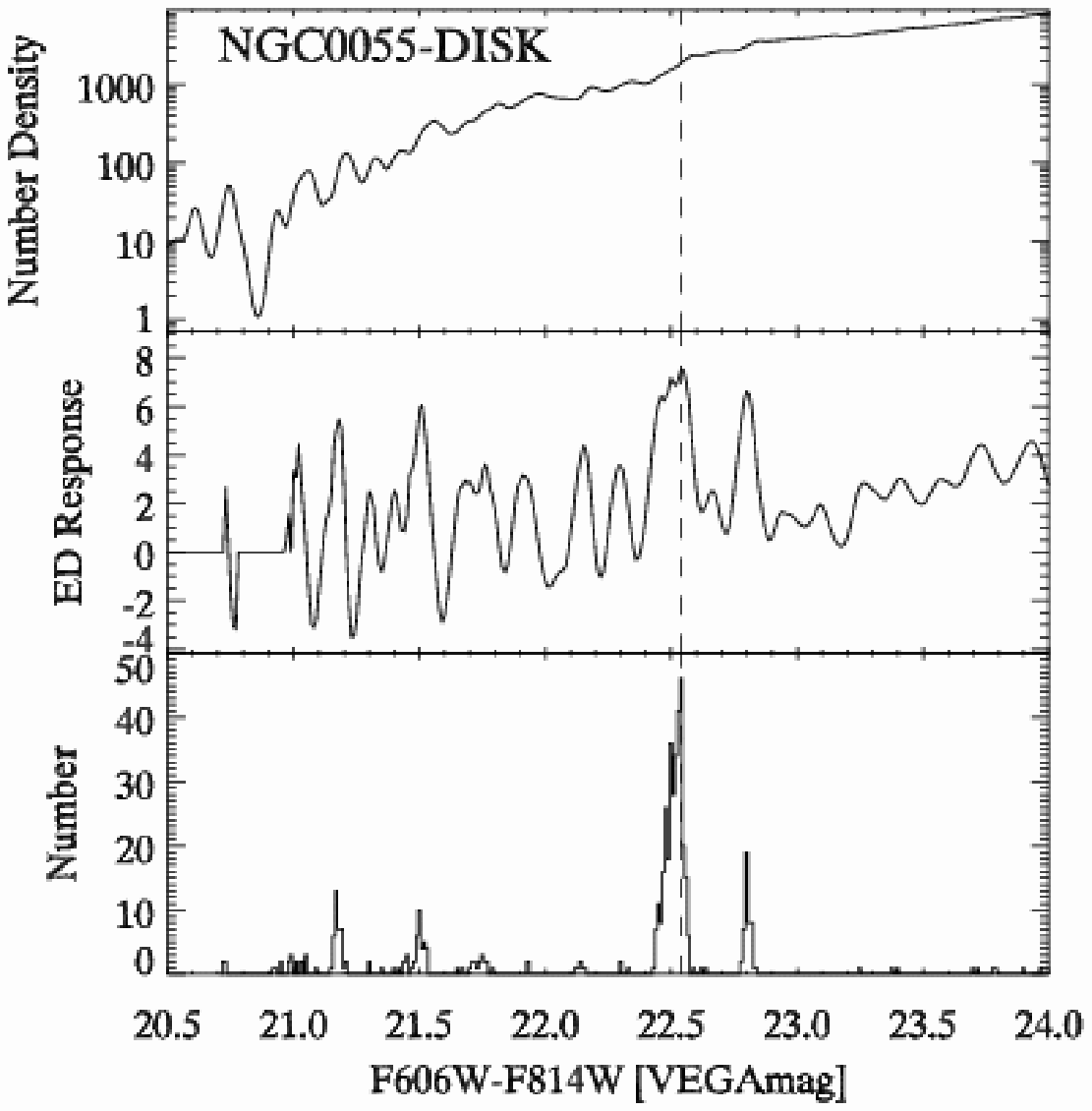}
\plottwo{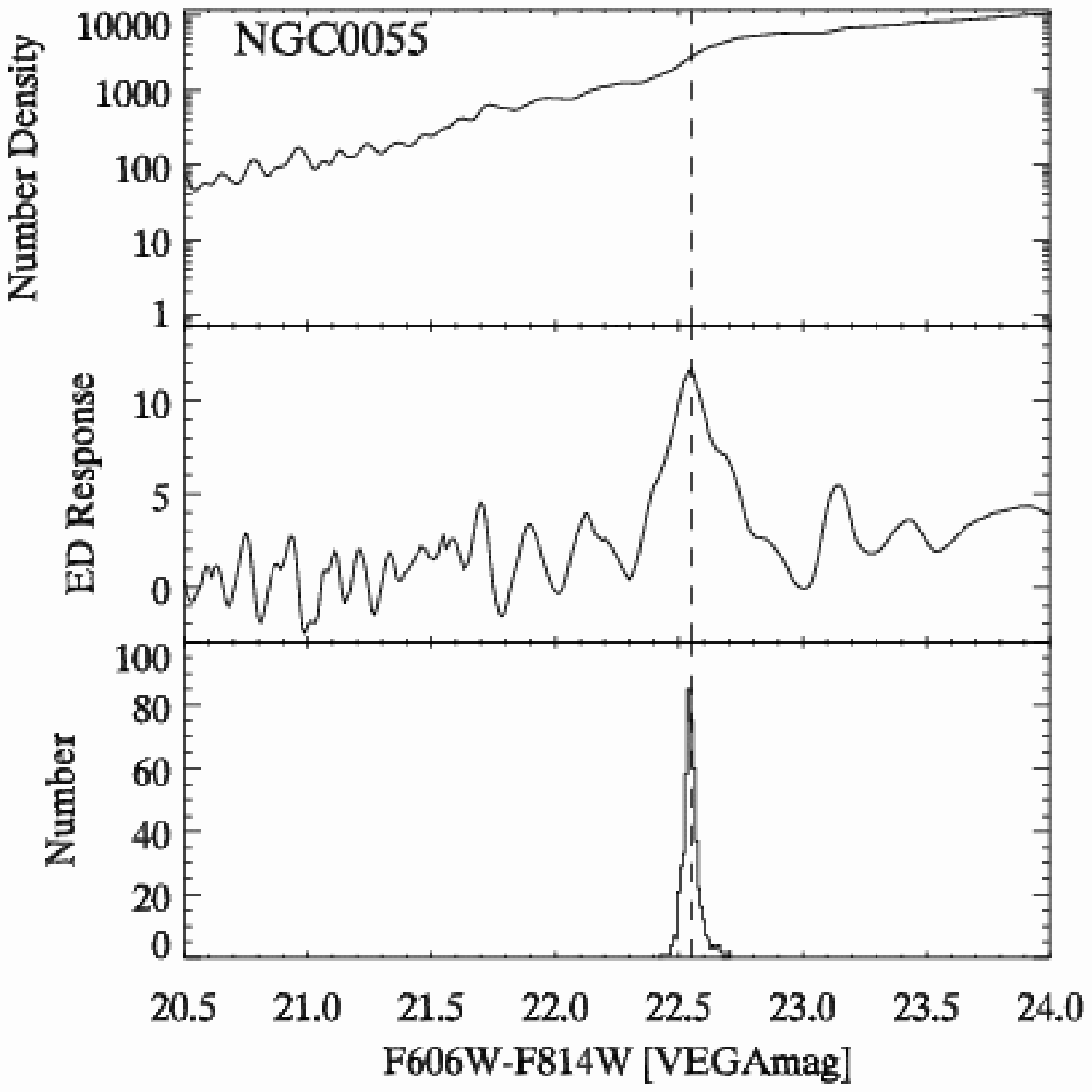}{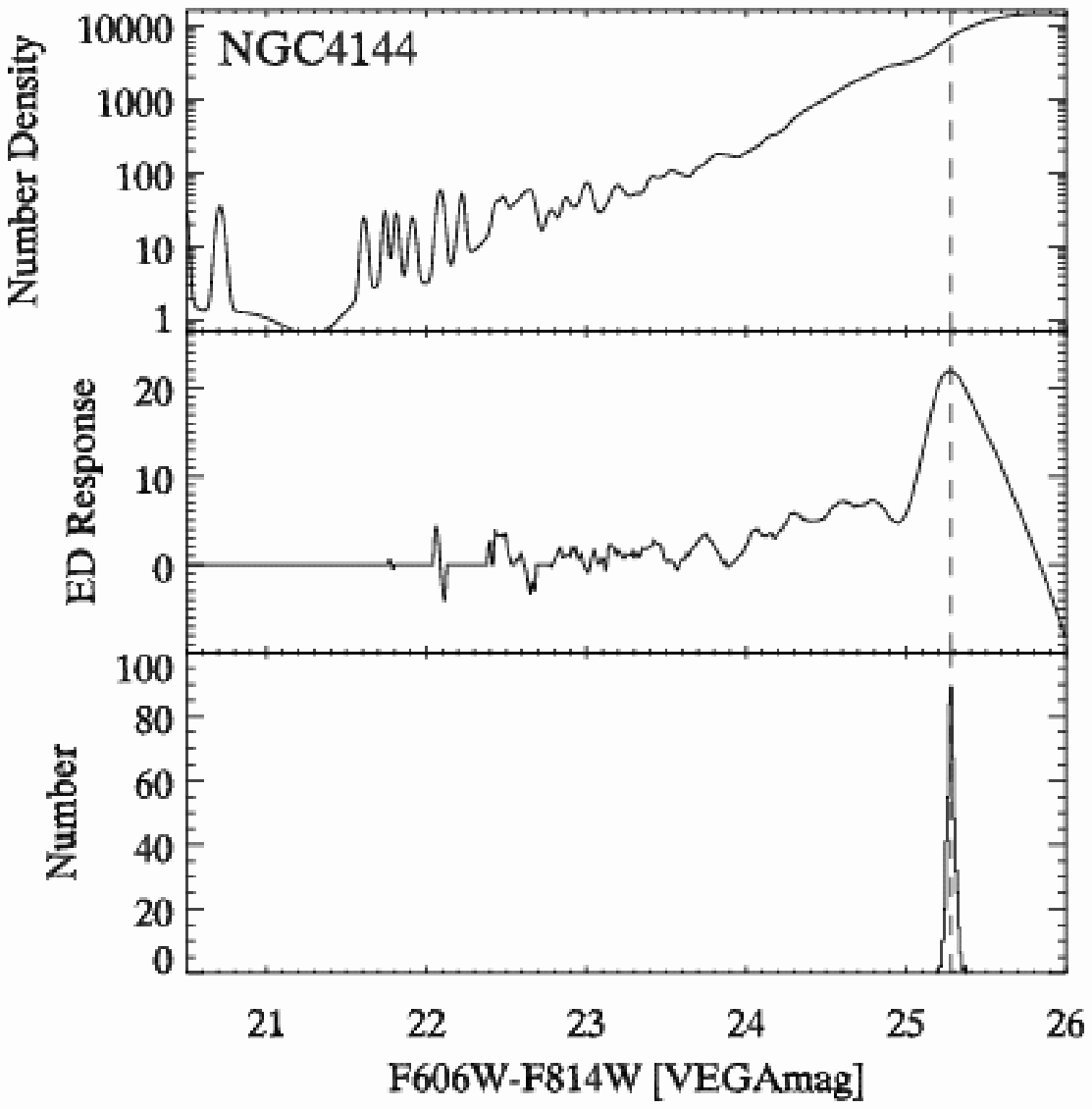}
\caption{TRGB determination for eight fields in six galaxies.  In each
  plot, the top panel shows the Gaussian-smoothed luminosity function as a
  function of the F814W magnitude, the middle panel shows the edge detection
  response to the luminosity function and the bottom panel shows a
  histogram of the edge-detection peak for 500 monte carlo
  simulations.  The 
  dashed line gives the position of the determined TRGB value. For
  more details see \S4.1.  }
\end{figure*}

\addtocounter{figure}{-1}
\begin{figure*}
\plottwo{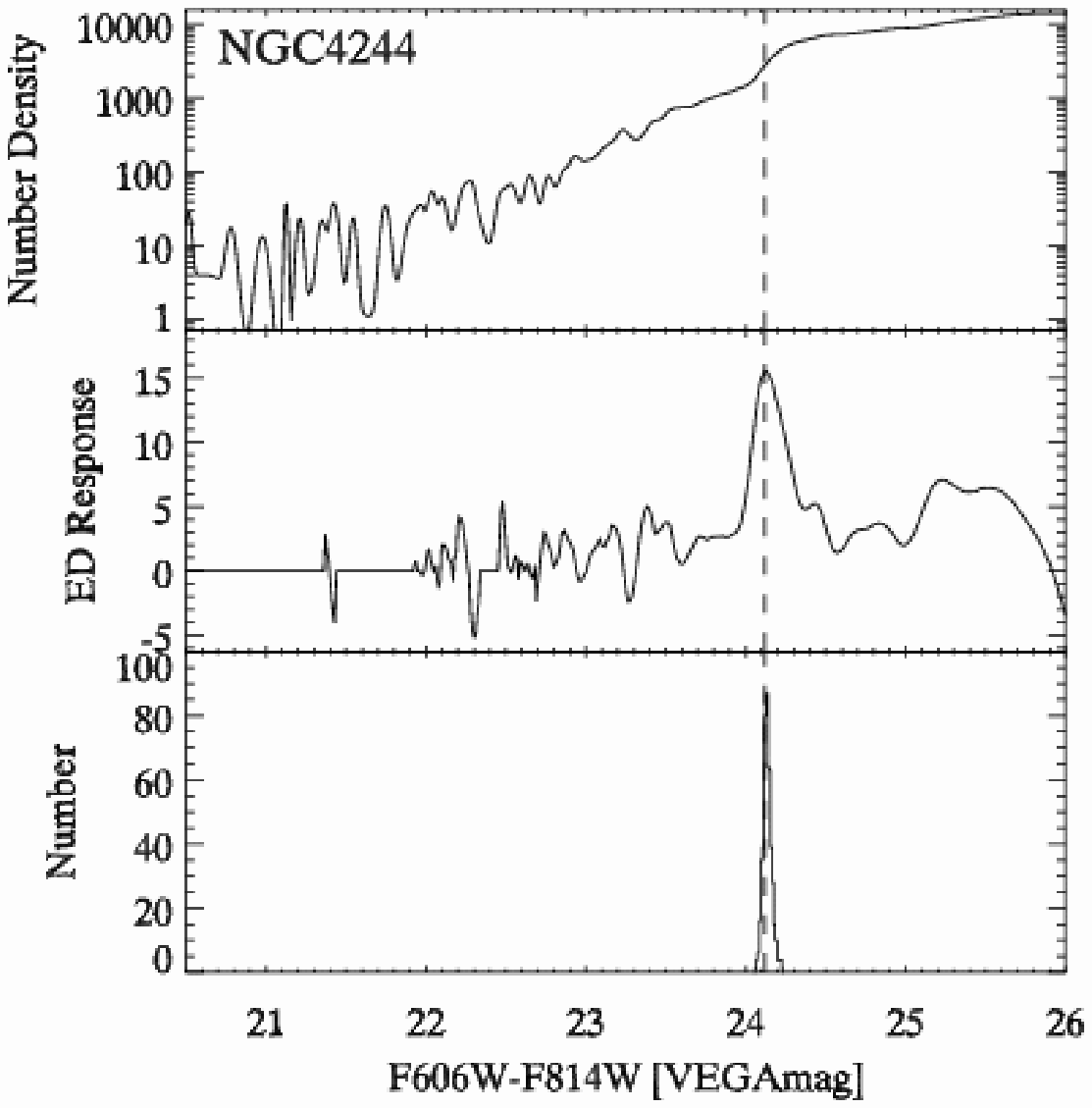}{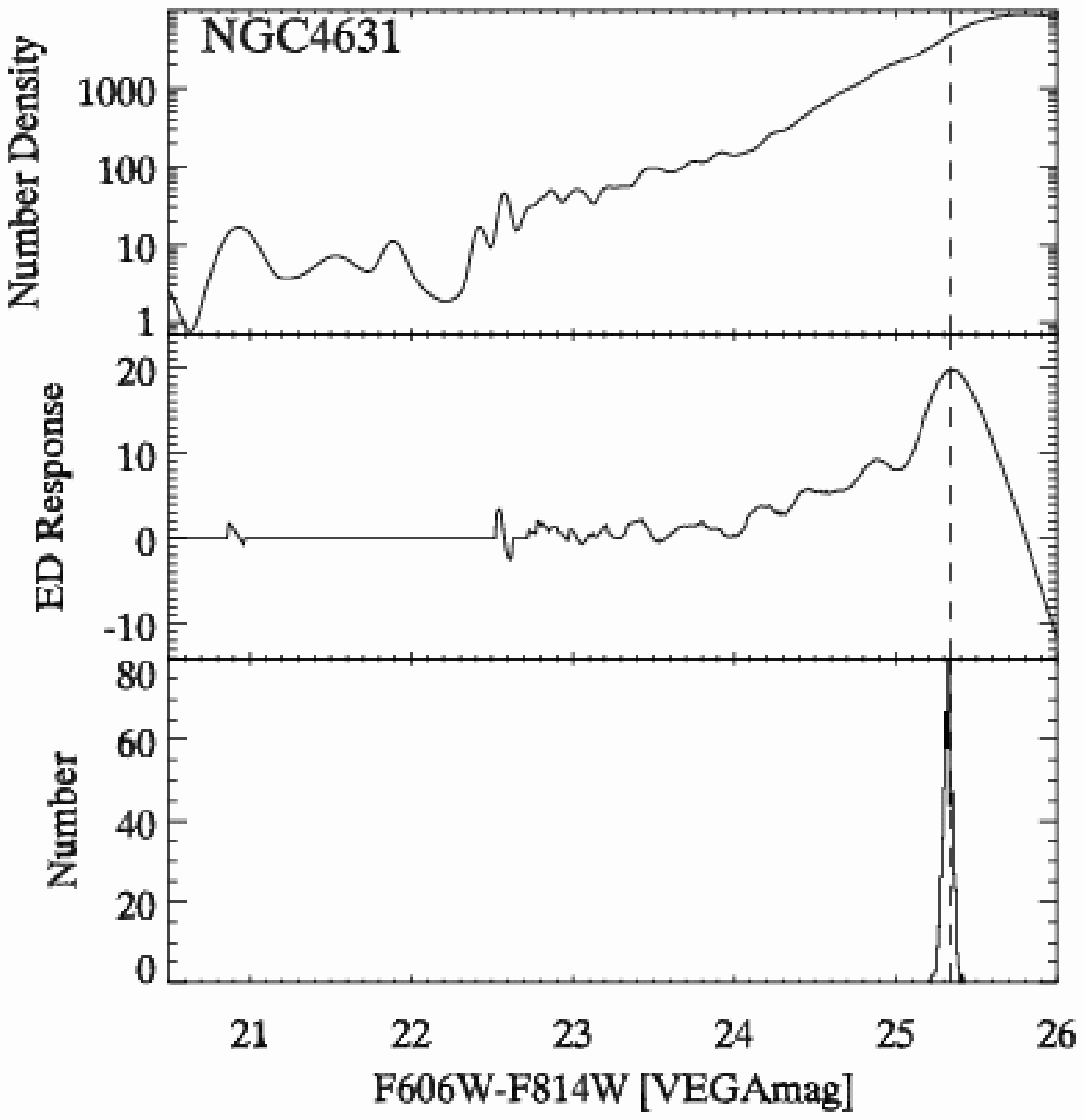}
\plottwo{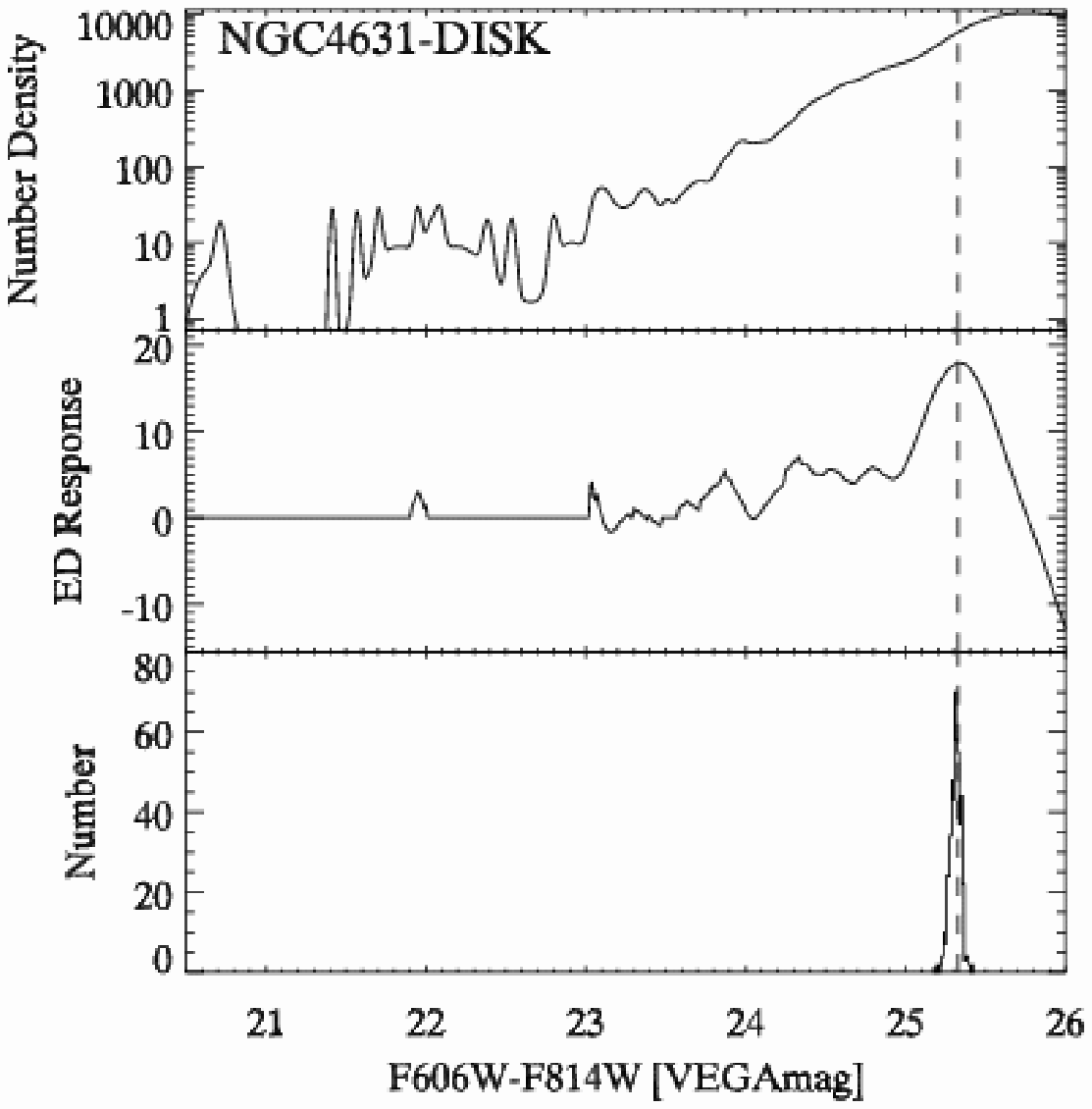}{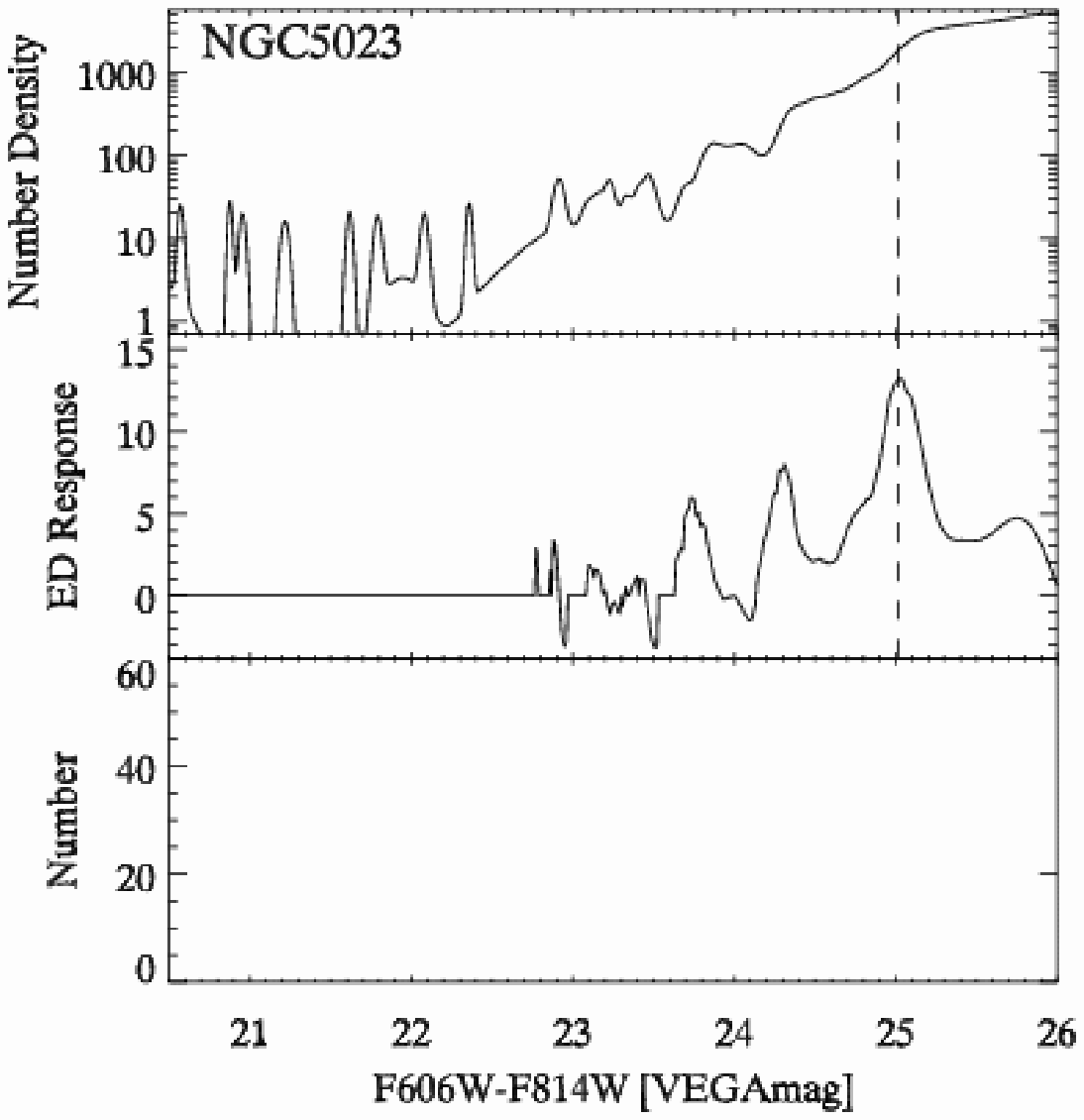}
\caption{{\it continued}}
\end{figure*}

\subsection{TRGB magnitudes}

The results of the TRGB determination for the eight fields in which
a clear RGB could be distinguished are shown in Table~5 and in the
CMDs in Fig.~6.  The F814W TRGB
magnitudes range from 22.54 for NGC 55 to 25.34 for NGC 4631.  The
uncertainties are typically $\sim$0.03 magnitudes, comparable to the errors
found using a similar method in \citet{mendez02} and to the systematic
errors in our magnitudes (\S2.3).  In all cases, the
value derived for the TRGB fell within one sigma of the Gaussian
fitted to the Monte Carlo tests.  The second column of Table~5 shows
the number of stars used in the TRGB determination of each galaxy.

\begin{deluxetable*}{lccccccc}
\tabletypesize{\scriptsize}
\tablecaption{TRGB results}
\tablehead{
     \colhead{Field}  &
     \colhead{\# of stars used} &
     \colhead{F814W$_{\rm TRGB}$} &
     \colhead{I$_{\rm TRGB}$} &
     \colhead{(F606W-F814W)$_{-3.5}$} &
     \colhead{(V-I)$_{-3.5}$} &
     \colhead{[Fe/H]} &
     \colhead{m-M$_{\rm TRGB}$} \\
     \colhead{}  &
     \colhead{} &
     \colhead{[VEGAmag]} &
     \colhead{[Johnson]} &
     \colhead{[VEGAmag]} &
     \colhead{[Johnson]} &
     \colhead{} &
     \colhead{} 
}

\startdata
IC 5052       & 18974 & 24.82$\pm$0.03 & 24.84$\pm$0.04 & 1.15 & 1.48 & -1.22 & 28.90 \\ 
NGC 55        & 53984 & 22.55$\pm$0.03 & 22.55$\pm$0.03 & 1.08 & 1.42 & -1.41 & 26.63 \\ 
NGC 55-DISK   & 56651 & 22.54$\pm$0.03 & 22.55$\pm$0.03 & 1.04 & 1.36 & -1.62 & 26.62 \\ 
NGC 4144      & 19132 & 25.28$\pm$0.02 & 25.29$\pm$0.03 & 1.06 & 1.39 & -1.50 & 29.36 \\ 
NGC 4244      & 27797 & 24.12$\pm$0.03 & 24.14$\pm$0.03 & 1.08 & 1.41 & -1.45 & 28.20 \\ 
NGC 4631      & 11204 & 25.34$\pm$0.03 & 25.37$\pm$0.03 & 1.13 & 1.47 & -1.25 & 29.42 \\ 
NGC 4631-DISK & 12058 & 25.33$\pm$0.03 & 25.36$\pm$0.04 & 1.06 & 1.38 & -1.54 & 29.41 \\ 
NGC 5023      &  8376 & 25.02$\pm$0.03 & 25.04$\pm$0.03 & 1.03 & 1.33 & -1.71 & 29.10 \\ 
\enddata			  	  
\end{deluxetable*}		  	  

For each of the galaxies shown in Table~5, a clear TRGB is seen
(Fig.~6).  For the other galaxies, we are limited in our  
ability to determine TRGB distances by incompletness.  The F814W 90\%
completeness level is 
$\sim$25.5 in uncrowded areas  at RGB colors and drops off quickly
towards fainter 
magnitudes.  The only two other possible detections are NGC 891 and
NGC 4517, both of which we measured to have TRGB magnitudes of 25.60.
However, examination of the CMDs reveals that it is impossible to
exclude the possibility that 
the stars below that magnitude are AGB or RHeB stars.  In addition,
the Monte Carlo tests find a 
wide range of values for the TRGB and their results are not 
consistent with the a TRGB magnitude of 25.60.  We suggest that the
derived TRGB magnitude for these galaxies is a lower limit.  
For the other 8 fields in our sample, there were typically too few
stars above 3$|$z$_{1/2}|$ to distinguish any red giant branch. 

\subsection{Distance Moduli}

The I-band absolute magnitude of the TRGB has been found to be at
M$_{\rm I}\sim-4$ via many methods 
\citep[e.g.][]{mendez02,lee93,ferrarese00}.  Because the photometric
system we use is similar to the Johnson system, we can expect that the
TRGB is at a similar F814W absolute magnitude (M$_{\rm F814W}$).  In order
to accurately determine the distance, we need to (a) determine the
observed TRGB magnitude in the Johnson I system, or (b) determine the
exact value M$_{\rm F814W}$ of the TRGB.  We take both of these
approaches below.  

Using the transformations provided to us by Sirianni, we transformed
the stellar data for each galaxy to the V and I filter and
redetermined the TRGB exactly the same as above, 
but using V-I cuts of 0.5 and 1.9.
The results are shown in Table~5.  By calibrating the TRGB with
Cepheid distances, \citet{ferrarese00}, found the TRGB
to be at M$_{\rm I} = -4.06 \pm 0.07$ (random) $\pm 0.13$
(systematic).  The magnitudes of I$_{\rm TRGB}$ 
agree with F814W$_{\rm TRGB}$ to within 0.03 magnitudes for each of
the 8 fields, but  on average I$_{\rm TRGB}$ is fainter by $\sim$0.02
magnitudes.  This suggests that the TRGB is at M$_{\rm F814W}\sim-4.08$.  

\begin{figure}
\plotone{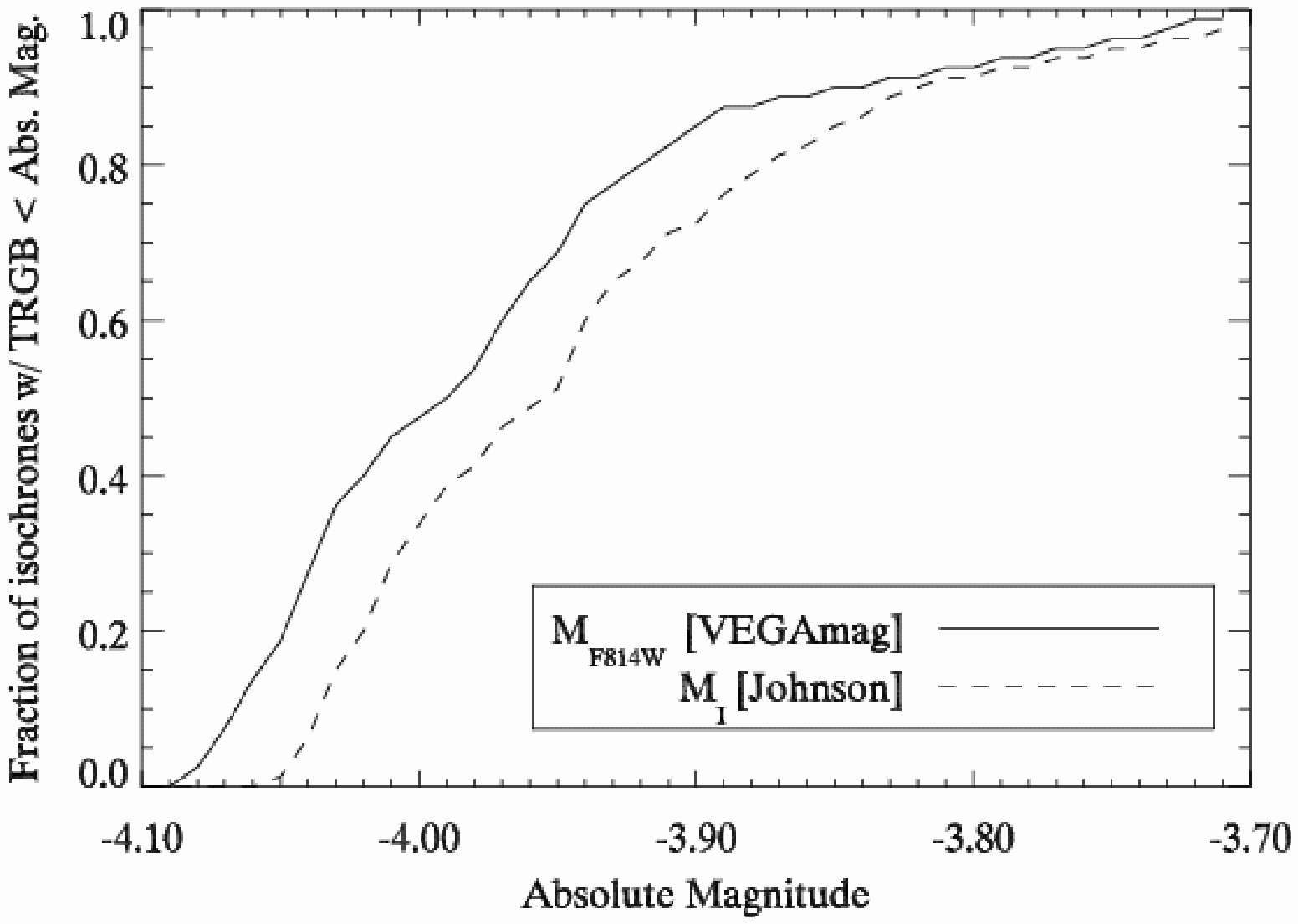}
\caption{Cumulative distribution of Padova isochrones with TRGB
  magnitudes brighter than a given absolute magnitude.  Solid line
  shows the distribution for M$_{\rm F814W}$ VEGAmag isochrones, while
  the dashed line shows the Johnson M$_{\rm I}$ values.}
\end{figure}

Alternately, we can use the theoretical Padova isochrones (see \S2.5),
which exist in both filter systems.  For
both sets of isochrones we take all tracks older than 2 Gyr and
with [Fe/H] between -2.3 and -0.7.  Figure~8 plots the cumulative
distribution of the TRGB
magnitude in both the F814W VEGAmag system (solid) and in Johnson I
(dashed).  A great majority of the tracks have TRGB values in both
systems between -3.8 and -4.1.  The mean, median and brightest values
of the absolute magnitude of the TRGB
are all $\sim$0.03 magnitudes brighter in F814W VEGAmag than in
Johnson I, in good agreement with the empirical calibration above.  We
therefore assume that 
M$_{\rm F814W} = -4.08$ for determining distances from the observed
TRGB magnitudes.  The resulting distance moduli
are shown in the final column of Table~5.  The errors in the distances
are the random error associated with the TRGB measurement added in
quadrature with the 0.07 mag error from the \citet{ferrarese00}
calibration, resulting in a $\sim$0.08 mag random error.  The
systematic error is approximately 0.15 magnitudes, slightly 
higher than the \citet{ferrarese00} value due to the
change in filter system.

Comparing these distances to the literature values compiled in Table~1
reveals good agreement 
on two galaxies, with less good agreement for the other four.
We note that our limit at F814W$_{\rm TRGB}$
of $\sim$25.5 gives a distance modulus limit of $\sim$29.6.  Based on
previous distance measurements, we do not expect to detect
the TRGB in any other galaxies other than those listed in Table~5 except
possibly for NGC 891 and NGC 4517, whose non-detection we discussed above.

We now compare our distances with previous distances on a case-by-case
basis.  The listed $\Delta$(m-M) value is found by taking our distance
modulus and subtracting the previous measurement.  
\begin{itemize}
\item IC 5052 ($\Delta$(m-M)=0.10):  The previous
  distance measurement is from the IR Tully-Fisher relation as given in
  \citet{bottinelli88}.  In fact, \citet{bottinelli88} gives two
  values for the distance modulus, calculated using different
  Tully-Fisher relations, the mean of which gives the distance modulus we
  have found.  
\item NGC 55 ($\Delta$(m-M)=0.35): The distance given in Table~1 is from
  \citet{karachentsev03a}, who reinterprets older HI Tully-Fisher
  data collected from various sources by \citet{puche88} to determine
  the distance of 1.8 Mpc (m-M=26.28).  Previous measurements using
  the magnitudes of 
  supergiants \citep{pritchet87} from ground-based narrow band imaging
  find an even closer distance of 1.38 Mpc (m-M=25.7) which disagrees
  even more severely with our measured distance.  
  We suggest that these previous measurements may be unreliable,
  especially the Tully-Fisher relationship in this irregular galaxy.
  We also note that our distance modulus agrees within the errors to
  the TRGB \& Cepheid distance to NGC 300
  \citep{butler04,freedman01}, another galaxy 
  in the Sculptor Group.  Assuming similar distance moduli,
  the separation of 8 degrees between these galaxies translates to a
  physical distance of $\sim$300 kpc.  
\item NGC 4144 ($\Delta$(m-M)=1.43): The Table~1 m-M=27.93 value is from
  the \citet{bottinelli88} IR Tully-Fisher relation.  Other estimates
  give wildly varying distances: 
  a previous B band Tully-Fisher distance \citep{bottinelli85} gave 
  m-M=28.87, while a brightest star measurement by
  \citet{karachentsev98} gives m-M=29.9.  Our distance measurement
  falls between these latter two.  
\item NGC 4244 ($\Delta$(m-M)=-0.06): The close agreement between our
  distance and the distance in Table~1 is unsurprising because both
  were derived using the TRGB method from HST data.  We note that
  a previous B-band Tully-Fisher measurement finds m-M=27.77
  \citep{bottinelli84}, $\sim$0.5 magnitudes away from the TRGB distance.
\item NGC 4631 ($\Delta$(m-M)=1.24): The previous measurement of
  m-M=28.18 is from the
  B-band Tully-Fisher relation of \citet{bottinelli85}.  Measurements of
  the HI and CO Tully-Fisher \citep{schoniger94} relation suggest an
  even closer distance modulus of 27.5 to 27.8, in even worse
  agreement with the TRGB distance.  We note that this
  galaxy is undergoing interactions \citep{rand94} and has nearby
  companions, both of which could make the Tully-Fisher distances less
  reliable.  
\item NGC 5023 ($\Delta$(m-M)=0.35):  This difference in m-M is within
  the very large error (0.5 mag) given by \citet{bottinelli85} for the B-band
  Tully-Fisher measurement of this galaxy.  
\end{itemize}

To provide another check on our distance measurements, we
compare our TRGB distances to the K band/V$_{\rm max}$ Tully-Fisher relation
presented in \citet{verheijen01}: 
\begin{equation}
{\rm
M_{K} = -1.62\pm0.49 - 8.5\pm0.2\:log(2 V_{max})
}
\end{equation}
Figure~9 shows the comparison between this Tully-Fisher relation and
the distances we have derived.  
The magnitudes shown were calculated using the 'total' 
K$_{\rm s}$ magnitude (K$_{\rm tot}$ in Table~1) from
\citet{jarrett03} and our TRGB distances.  
The V$_{\rm max}$ values are those shown in Table~1 \citep{paturel03}
and represent the maximum circular velocity in that galaxy.  We note
that K$'$ filter used for deriving the Tully-Fisher relationship has a
similar bandpass to the K$_{\rm s}$ filter used in 2MASS, and so any
differences between the magnitude systems should be small.  For the six
galaxies that we have derived distances to, the scatter around the
Tully-Fisher relation is 0.65 magnitudes.  This is similar to the
0.71 magnitude scatter around the relation found by
\citet{verheijen01}.

\begin{figure}
\plotone{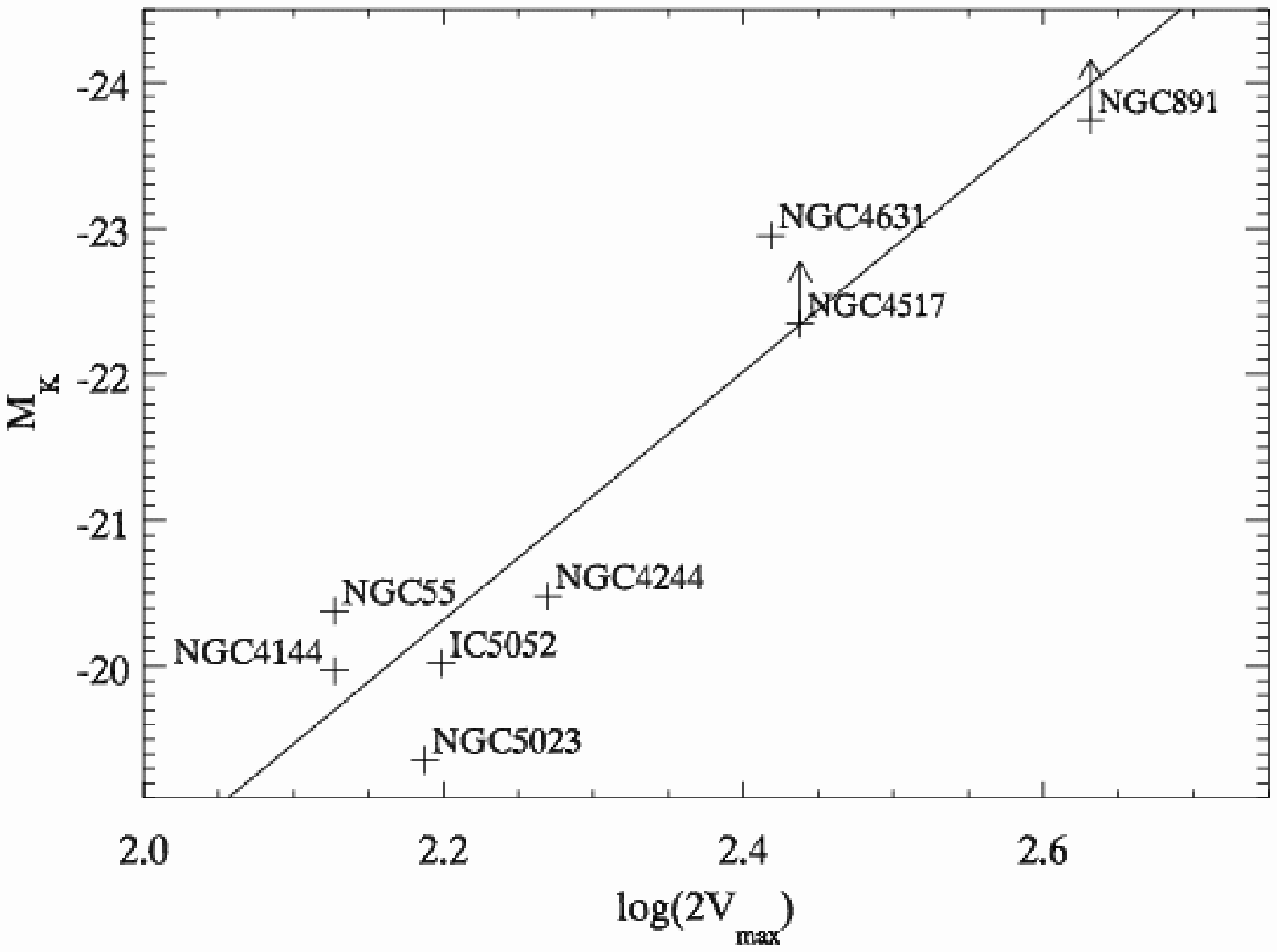}
\caption{A comparison of the derived TRGB distances (data points)
  vs. the K band Tully-Fisher relation from \citet{verheijen01}
  (line).  In addition to the six galaxies with detected distances, we
  show NGC 891 and NGC 4517 assuming a lower limit to the distance
  modulus of 29.68.}
\end{figure}

Overall, we find that there are plausible reasons for the differences
between our distances and those previously measured.  
The agreement found between different fields of the same galaxy as
well as the consistency of our TRGB distances with the K band
Tully-Fisher relationship and the previous TRGB measurement (NGC
4244) all give us reason to believe the distances
given here are more accurate than previous distances found in the
literature.  We also note that the relative distances between the
galaxies should be accurate to within the $\sim$0.03 magnitude errors
in the TRGB determination.  

\subsection{Metallicity of the RGB}

Table~5 shows the color of the RGB at an absolute magnitude of -3.5
in both the Johnson system, (V-I)$_{-3.5}$ and the ACS VEGAmag system,
(F606W-F814W)$_{-3.5}$.  These values were found by fitting a
Gaussian to the distribution of stars within an 0.2 magnitude band
centered at an absolute magnitude of -3.5 (assuming TRGB at 
M$_{\rm F814W}$=-4.08 and M$_{\rm I}$=-4.06) .  The value for
(V-I)$_{-3.5}$ has  
been used by \citet{lee93} to derive the metallicity as follows:
\begin{equation}
{\rm
[Fe/H] = -12.64 + 12.6(V-I)_{-3.5} - 3.3(V-I)^{2}_{-3.5}
}
\end{equation}
The [Fe/H] values from this equation are shown in Table~5 and range
from -1.2 to -1.7.  Comparison
of the (F606W-F814W)$_{-3.5}$ values (Table~1) with Padova isochrones
shows similar results, with the data
falling just to the metal rich side of the [Fe/H]=-1.3 isochrone.
This confirms that the population of stars we are
using for TRGB distance determination are metal poor and suggests that
they are part of some kind of halo or thick disk system.  Note also
that the outer fields of NGC 55 and NGC 4631 have lower metallicities
than the inner fields as expected in some disk formation models.  We
will explore these trends more fully in Paper II.

\section{Summary}

In this paper we have presented the basic properties and
data reduction of a sample of edge-on galaxies
observed with ACS camera on HST.  We summarize our findings below:
\begin{itemize}
\item A number of the sample galaxies are close enough to be partially
  resolved into stars with an identifiable Red Giant Branch,
  Asymptotic Giant Branch, Main Sequence, and Helium Burning sequences.  
\item We identify a candidate dwarf galaxy near NGC 4631 that has
  a mass, age, and size similar to an OB association, but lies 2 kpc
  above the disk.  We also find LMC stars in the IRAS 06070-6147
  field, $\sim$9$^\circ$ from the LMC center.   
\item We determine scale lengths and heights for each galaxy using
  single-disk fits to K-band 2MASS images.
\item TRGB distances are derived to eight fields in six galaxies.
  These are found to have distance moduli between 26.62 and 29.42.
\item The metallicities of the high latitude stars appear to be
  metal-poor with [Fe/H] $<$ -1.  
\end{itemize}

This work is the first in a series of papers, the next of which 
focuses on the stellar populations of the galaxies as a function of
scale height (Paper II).  Future work includes a third paper
on the star cluster content of the sample galaxies.

The authors would like to acknowledge the help of Tom Jarrett
in obtaining 2MASS data, Thomas Brown and Marco Sirianni for their help in
obtaining ACS photometric calibrations and Leo Girardi for providing
isochrones in the ACS filter system.  Additionally we'd like to thank
Simone Bianchi, Paul Hodge, Kevin Covey, Vandana Desai, Peter
Yoachim, and our referee for helpful discussions and
comments. Julianne J. Dalcanton was partially supported by the 
Alfred P. Sloan research foundation.  This research has made use of
the NASA/IPAC Extragalactic Database (NED) which is operated by the
Jet Propulsion Laboratory, California Institute of Technology, under
contract with the National Aeronautics and Space Administration.  
This work was supported by HST-GO-09765.

\appendix

\section{K band disk fits}

We describe here the fits performed on 2MASS K band images to
determine a consistent set of structural properties in our galaxies.
We obtained 2MASS K band images of each galaxy in our sample from the 
Large Galaxy Atlas \citep{jarrett03}.  Only one of the galaxies, NGC
3501, was not in the Large Galaxy Atlas, and this we obtained via the
2MASS Extended Source Image
server\footnote{http://irsa.ipac.caltech.edu/applications/2MASS/PubGalPS/}.  

We fit these K band images to an edge-on single disk model from
\citet{vanderkruit81}: 
\begin{equation}
  \Sigma(R,z)=\Sigma_{0} (R/h) K_{1}(R/h_{R}) sech^{2}(z/z_{0})
\end{equation}
where $R$ and $z$ are the radial and vertical coordinates, $\Sigma_0$ is
the edge-on central surface brightness, $h_{R}$ is the radial scale length and
z$_{0}$ is the vertical scale height of an isothermal stellar
population.  Note that z$_0$ is 
different from an exponential scale height.  For this reason we use
the z$_{1/2}$ parameter, which gives the height above and below the
plane that contains 50\% of the K band light to facilitate comparison
with other studies.  For our adopted profile in Eq.~1, z$_{1/2}$=0.549z$_0$.

Before fitting, the 2MASS images were cropped and sources
were identified using SExtractor \citep{bertin96}.   Sources other
than the target galaxy were masked and the masked image was fit to the
single disk model of Eq.~1 using a Levenberg-Marquardt
least-squares fit with a uniform weighting of all the unmasked pixels.
For galaxies with obvious bulge components (IRAS 06070-6147, NGC 891,
NGC 3501, NGC 4206, NGC 4244, and NGC 4517), we masked out the central
portion of the galaxy with a radius between 5 and 50 arcsec depending
on the size of the bulge.  We attempted to fit a second disk component
to the galaxies, but found that it did not increase the quality of the
fit.  Four galaxies, NGC 55, NGC 4183 and NGC 4631, showed significant
asymmetry complicating the fitting process, while one galaxy, NGC 891
had a significant dust lane even at K band.  The formal errors on
z$_0$ and $h_R$ were at most 4\%, however, the actual errors will in
most cases be dominated by source structure in the galaxies.

The results of the single disk fits are shown in Table~1 (RA, Dec and
position angle, given E of N) and Table~2 ($\Sigma_0$, $h_{R}$, and z$_0$).
Also shown in Table~2 are the ratio of scale length to height
$h_{R}$/z$_0$, the half-light height z$_{1/2}$ and the $h_{R}$ and z$_0$
values in parsecs assuming the distances in Tables 1 \& 5 (where
available, TRGB distances from Table~5 are used).
The values for the scale length, $h_{R}$, range between 16$\arcsec$ and
91\arcsec, and in physical coordinates between 0.9 and 7 kpc.  For
z$_0$, the range is 2$\arcsec$ to 24$\arcsec$ and 300 to 1100 pc.  For
comparison the Milky Way has $h \sim 3$ kpc \citep{klypin02} and
z$_{0} \sim 700$ pc \citep{vanderkruit81}.  The typical $h_{R}$/z$_0$ values
for our galaxies are $\sim$4, similar in value to the Milky Way and
the 31 edge-on galaxies presented in \citet{pohlen00}.

\end{document}